\documentclass[aip,
 amsmath,amssymb,
 reprint,%
]{revtex4-1}

\usepackage{graphicx}
\usepackage{dcolumn}
\usepackage{bm}

\usepackage[utf8]{inputenc}
\usepackage[T1]{fontenc}
\usepackage{mathptmx}
\usepackage{etoolbox}
\usepackage{amssymb}
\usepackage{amsmath}
\usepackage{amsthm}
\usepackage{natbib}

\newtheorem{theorem}{Theorem}

\newtheorem{lemma}{Lemma}

\makeatletter
\def\@email#1#2{%
 \endgroup
 \patchcmd{\titleblock@produce}
  {\frontmatter@RRAPformat}
  {\frontmatter@RRAPformat{\produce@RRAP{*#1\href{mailto:#2}{#2}}}\frontmatter@RRAPformat}
  {}{}
}%
\makeatother
\begin{document}

\preprint{AIP/123-QED}

\title[Global synchronization theorem for coupled swarmalators]{Global synchronization theorem for coupled swarmalators}
\author{Kevin O'Keeffe}
 \altaffiliation[]{Starling Research Institute}

\date{\today}

\begin{abstract}
The global stability of oscillator networks has attracted much recent attention. Ordinarily, the oscillators in such studies are motionless; their spatial degrees of freedom are either ignored (e.g. mean field models) or inactive (e.g geometrically embedded networks like lattices). Yet many real-world oscillators are mobile, moving around in space as they synchronize in time. Here we prove a global synchronization theorem for such swarmalators for a simple model where the units' movements are confined to a 1d ring. This can be thought of as a generalization from oscillators connected on random networks to oscillators connected on temporal networks, where the edges are determined by the oscillators' movements.
\end{abstract}

\maketitle

\begin{quotation}
An interplay between swarming (self-organization in space) and synchronization (self-organization in time) occurs all over Nature and technology, from starfish embryos \cite{tan2022odd} and vinegar eels \cite{quillen2021metachronal} to magnetic domain walls \cite{hrabec2018velocity} and Quincke rollers \cite{liu2021activity}. Yet we know surprisingly little about this dual form of emergence from a theoretical point of view. Even basic questions, like what kinds of collective states emerge and what determines their stabilities and bifurcations, have only recently begun to be addressed \cite{tanaka2007general,iwasa2012various,o2017oscillators,adorjani2024motility,kongni2023phase,degond2022topological,degond2023topological,gong2024approximating,smith2024swarmalators,levis2019activity}. This paper contributes to the effort to understand such swarmalators theoretically. We prove a global synchronization theorem for a simple model of swarmalators which run on a 1d ring. The theorem pinpoints the conditions under which perfect synchrony -- all swarmalators having the same phases and positions -- occurs with probability $1$, for all random initial conditions except for a set of measure zero. Our proof hold for any finite population size $N$ and exploits the fact that the model is a \textit{transformed} gradient system, a structure that is often overlooked in dynamical system studies and thus may be of general interest. The proof also has a microstate/macrostate flavour that may appeal to statistical physicists and related researchers. 
\end{quotation}

\section{Introduction}
A fascinating class of problems in applied science today involves the fusion of two types of self-organization: swarming, where the individual units coordinate their spatial positions, and synchronization, where the units coordinate the timing of an internal oscillation.

Take colloidal micromotors which are envisioned to revolutionize nanotechnology and related disciplines \cite{wang2015one,fernandez2020recent}. When subject to magnetic fields, the individual units' dipole vectors begin to oscillate and synchronize \cite{yan2012linking}. This sync process couples to the units' movements and thus can be used to influence how they cluster in space \cite{yan2012linking}. So-called ``sync-selected self-assembly" has great applied power. 
It has been used to degrade pollutants \cite{ursobreaking,dai2021solution,vikrant2021metal,tesavr2022autonomous}, repair electrical circuits \cite{li2015self}, and to shatter blood clots \cite{cheng2014acceleration,manamanchaiyaporn2021molecular}.

Spintronics, which aims to use magnetic domain walls as the information carriers for next generation memory devices \cite{linder2014superconducting}, also makes use of sync-selected self-assembly. By driving the walls’ dipole vectors into synchrony you can control their motion \cite{hrabec2018velocity} allowing you to do the logic required for memory storage. Swarm robotics is another example where robots programmed with synchronizable internal states are capable of diverse swarm-formations.




This usage of synchronization presents the sync community with an opportunity. If we could figure out how to apply our theories about regular (unmoving) oscillators to oscillators which are mobile, we could help usher in the bright future in nanotechnology and other fields that the mentioned applications promise to unlock. 

A natural place to start is to take a simple m oscillators and allow the oscillator to move around in the line or plane. A new stream of work \cite{tanaka2007general, iwasa2010dimensionality, o2017oscillators, schilcher2021swarmalators, kongni2023phase, yadav2024exotic, lizarraga2020synchronization, ha2019emergent, ha2021mean, gong2024approximating, adorjani2024motility}
has taken this approach and dubbed the entities swarmalators, since they generalize swarms and oscillators \cite{o2017oscillators}.

Theoretical results about swarmalators moving in 2d or 3d are few \cite{o2017oscillators,o2018ring,gong2024approximating,o2023solvable,smith2024swarmalators,lizarraga2024order}. By retreating to a 1d periodic domain \cite{o2022collective}, however, researchers have made more progress. Exact results for bifurcations have here been derived for swarmalators with heterogenous natural frequencies \cite{yoon2022sync}, Gaussian noise \cite{hong2023swarmalators}, and other effects \cite{anwar2024collective,lizarraga2023synchronization,sar2023swarmalators,o2022swarmalators,hao2023attractive,anwar2024forced,o2024stability,anwar2024forced,sar2024swarmalators}.

Yet even within the simplified setting of the 1d swarmalator model one class of results has yet to be addressed: global stability. That is the focus of this paper. 

Global stability is vital in applied settings. Consider the targeted drug delivery promised by micromotors. Instead of simply releasing the tumor-killing drug into the bloodstream and allowing a fraction to be absorbed at the tumor site, as is currently done, the idea here is to assemble `colloidal cages' that contain the drug until it reaches its destination. This is obviously far safer, with less toxic spillover to non-target sites during the drug's journey through the bloodstream.

Global stability, and not just local stability, of the colloidal cages is important here. If a closed sphere state is (locally) bistable with an open hemisphere state, say, environmental noise can knock the system from one state into the other, thereby triggering an unwanted release. To avoid this malfunction we need to know the parameter regions where the sphere state is globally attracting.

The same issues arise in swarm robotics, spintronics and related contexts where reliably and robustly realizing a single, desired formation is essential for applications. 

So a big open problem for the above applications -- and a natural entrypoint for sync researchers to make a contribution -- is to come up theoretical guarantees for the global stability of swarmalator populations.

This paper is a first case-study in this direction. We take the 1d swarmalator model \cite{o2022collective} and prove a theorem for when it globally synchronizes. This 1d model is one of the few models of swarmalators that is tractable, making it an ideal toy model for the micromotors, robotic swarms, and magnetic domain walls. It is also a clean generalization of the Kuramoto model, and so connects our work to a recent flurry of papers about the global synchronization of Kuramoto oscillators on random graphs \cite{lu2020synchronization,ling2019landscape,taylor2012there,kassabov2022global,abdalla2024guarantees,nagpal2024synchronization,harrington2023kuramoto,abdalla2024guarantees}. You can think of our work as a generalization from graphs with random edges to graphs with \textit{temporal} edges, determined by the swarmalators' (dynamic) positions in space.


\section{Model}  
The 1d swarmalator model\cite{o2022collective,yoon2022sync} is
\begin{align}
\dot{x}_i = \omega_i' +  \frac{J'}{N} \sum_j \sin(x_j - x_i) \cos(\theta_j - \theta_i) \\
\dot{\theta}_i = \nu_i' + \frac{K'}{N} \sum_j \sin(\theta_j - \theta_i) \cos(x_j - x_i)
\end{align}
where $x_i \in S^1$ is position of the $i$-th swarmalator on the unit circle and $\theta_i \in S^1$ is the phase. $(\omega', \nu')$ and $J',K'$ are the associated natural frequencies and couplings. This model has been used to model magnetic domain walls and other forms of driven matter \cite{o2023solvable,sar2023pinning}. It can also be thought of as the rotational part of the 2d swarmalator model \cite{o2017oscillators}. 

What's nice about the model is that if you switch to sum/difference coordinates $(\xi, \eta) := (x+\theta, x-\theta)$  a clean pair of coupled Kuramoto models falls out
\begin{align}
\dot{\xi}_i = \omega_i+  K r \sin(\phi - \xi) + J s \sin(\psi - \eta)  \\
\dot{\eta}_i =  \nu_i +  J r \sin(\phi - \xi) + K s \sin(\psi - \eta)  
\end{align}
where
\begin{align}
& Z_{\xi} := r e^{i \phi}= \langle e^{i \xi} \rangle \\
& Z_{\eta} := s e^{i \psi}  = \langle e^{i \eta} \rangle  \\
& (\omega, \nu) = (\omega' + \nu', \omega' - \nu') \\
& (J,K) = ( (J'+K')/2, (J'-K')/2)
\end{align}
The $r,s$ are the rainbow order parameters (so-called since they are maximal in the rainbow-like static phase wave state of the 2d model; see Fig2(c) in \cite{o2017oscillators}) and are nice generalization of the Kuramoto sync order parameter $R := \langle e^{i \theta} \rangle$.
\begin{figure}[t!]
\centering
\includegraphics[width = \columnwidth]{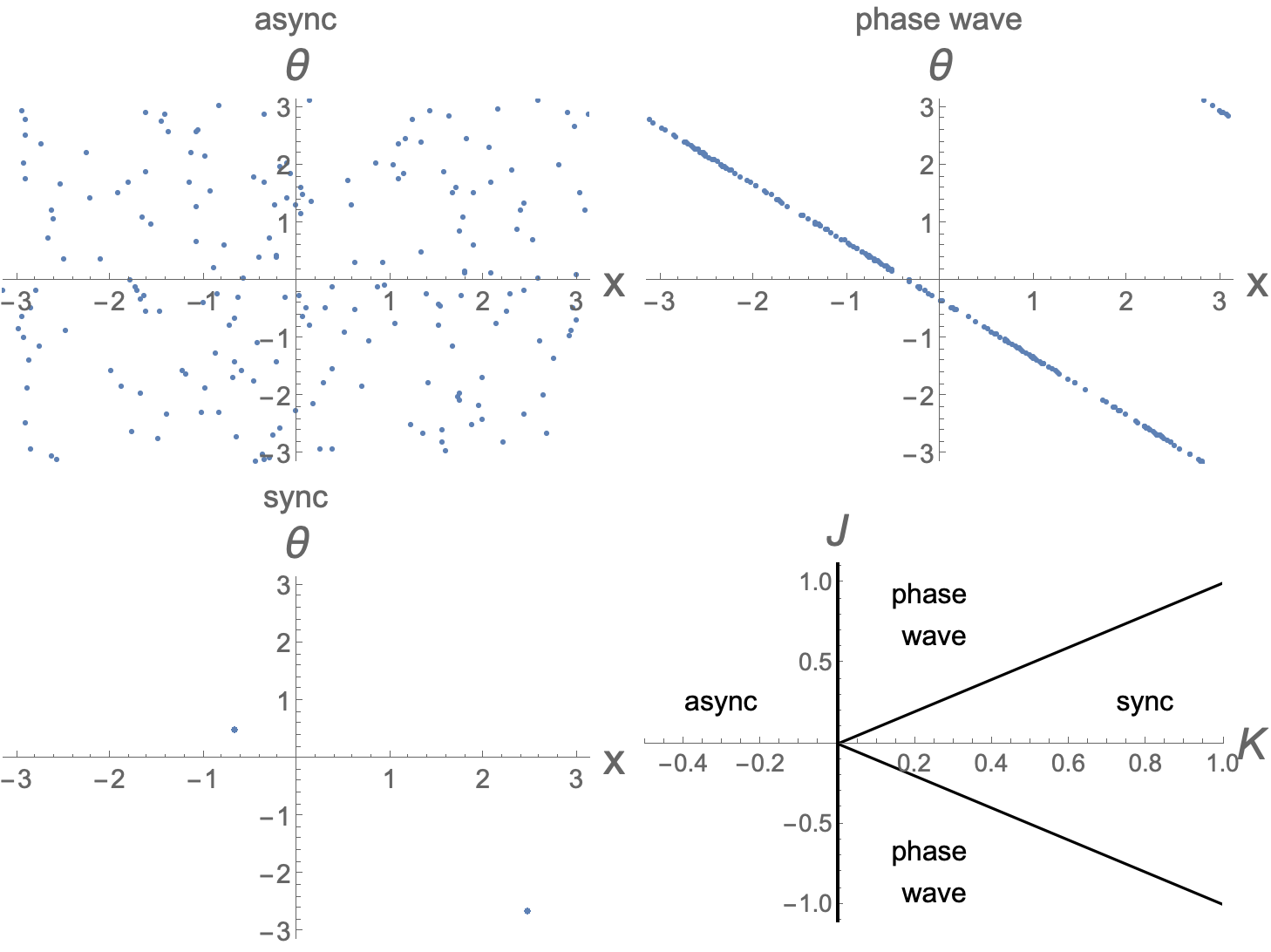}
\caption{(a)-(c) Collective states of 1d swarmalator model $(J', K') = (1,-2), (1,-0.5, 1,2))$ respectively. Here $(dt,T,N) = (0.1,100,200)$ and initial condition were drawn uniformly at random from $[-\pi,\pi]^2$. (d) Stability diagram in the $(J,K)$ plane valid in the continuum limit}
\label{states-identical}
\end{figure}

We are here concerned with the homogenous model where $(\omega_i, \nu_i) = (\omega, \nu) = (0, 0)$, where the setting to zero is valid via a change of frame. Numerics indicate its has three kinds of collective state for large $N$.  Figure~\ref{states-identical}(a)-(c) plots these $(x,\theta)$ plane. The states may be classified by the order parameter tuple $(r,s)$ like so
\begin{enumerate}
    \item \textit{Async} or $(r,s) = (0,0)$ state, swarmalators are distributed uniformly on $(-\pi,\pi)$ in both position and phase. There is no overall order.
    \item \textit{Sync} or $(r,s) = (1,1)$ state, swarmalators are fully sync'd in both $x$ and $\theta$. For most initial conditions two sync dots are formed, as you can see in Figure~\ref{states-identical}(b). These result from a $\pi$-symmetry in the governing ODEs; the transformation $(x,\theta) \rightarrow (x+\pi, \theta+\pi)$ leaves them unchanged. In $(\xi,\eta)$ coordinates however this manifests as a single dot. 
    \item \textit{Phase wave} or $(r,s) = (1,0), (0,1)$ state, where the swarmalators positions and phases are perfectly correlated $x_i = \pm \theta_i + C$, where the $\pm$ occur with equal probability. Put another way, the swarmalators are perfectly sync'd in either the $\xi$ or $\eta$ direction, and fully desynchronized in the other.
\end{enumerate}
All three states are \textit{macrostates}, in the sense they are defined by the the values of the (macroscopic) order parameters $(r,s)$. Each macrostate is composed of a family of microstates. For instance the async state corresponds to all configuration $z := (\xi_1, \dots \xi_N, \eta_1, \dots, \eta_N) $ such that
\begin{align}
    Z_{\xi} = N^{-1} \sum_j \cos \xi_j + \mathcal{I} \sin \eta_j = 0 \\
    Z_{\eta} = N^{-1} \sum_j \cos \xi_j + \mathcal{I} \sin \eta_j = 0 
\end{align}
where $\mathcal{I} = \sqrt{-1}$. More formally, the conditions above define a manifold in phase space we call the \textit{async manifold} $M_{async}$. The dimension of this manifold is $2N - 4$, which come from putting the $4$ constraints above (require the real and imaginary parts of both $Z_{\xi}$ and $Z_{\eta}$ to be zero) on the $2N$ dimensional phase space ($2N$ dimensional because there are $N$ swarmalator with $2$ coordinates $x,\theta$ each). 

The phase wave manifold $M_{wave}$ (defined by the two branches $(r,0) = (0,1)$ and $(r,s) = (1,0)$) has dimension $N-1$. Consider the $(r,s) = (0,1)$ branch. The $r=0$ piece has $2$ constraints (same as the async manifold; both real and imaginary parts have to be zero). The $s=1$ piece, which signifies full sync $\eta_i = C_1$, has $N-1$ constraints (since $C_1$ can be arbitrary). Subtracting these from the $2N$ phase space gives $\dim M_{wave} = N-1$.

Finally, the sync manifold defined by $r=s=1$, equivalent to $\xi_i = C_1, \eta_i = C_2$ has dimension $2$ for the reason described above. Collecting all these:
\begin{align}
    & \dim M_{async} = 2N-4 \label{dim_async}  \\
    & \dim M_{wave} = N-1 \label{dim_wave} \\
    & \dim M_{sync} = 1 \label{dim_sync}
\end{align}
These will become important later. These dimensions may be related to the entropy of each macrostate, if one is interested in a statistical mechanics interpretation.

\begin{figure}[t!]
\centering
\includegraphics[width = \columnwidth]{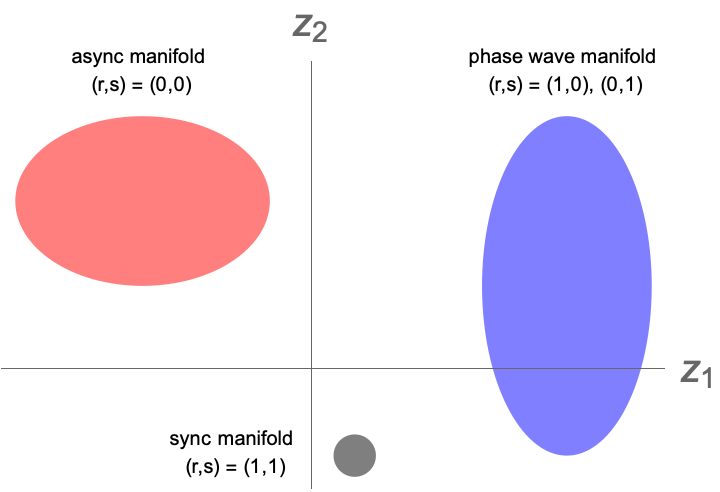}
\caption{Schematic of phase space for the 1d swarmalator model showing the sync, phase wave, and async manifolds. Lemma 2 proves  that all fixed points of the model lie on one of these manifolds expect for the special case $J = \pm K$. Here $z_i := (x_i, \theta_i)$ are the state variables and only a 2D slice is shown. See text.}
\label{schematic}
\end{figure}

Figure~\ref{schematic} gives a simplified schematic of these three manifolds in phase space $z := (x_1, \theta_1, x_2, \theta_2, \dots)$ where $z_i = (x_i, \theta_i)$. The schematic is not at all geometrically accurate. We give it just so you have a working mental model for how the phase space is partitioned up.

Getting back to the model's states, Figure~\ref{states-identical}(d) shows where each state occurs in the $(J,K)$ plane. Previous work derived the threshold boundaries by using linearization to find each state's local stability \cite{o2022collective}. Our aim here is to find the global stability of the sync state (the global stability of the async and phase wave are out of scope).

\section{Sketch of proof}
\begin{figure}[t!]
\centering
\includegraphics[width = 0.9\columnwidth]{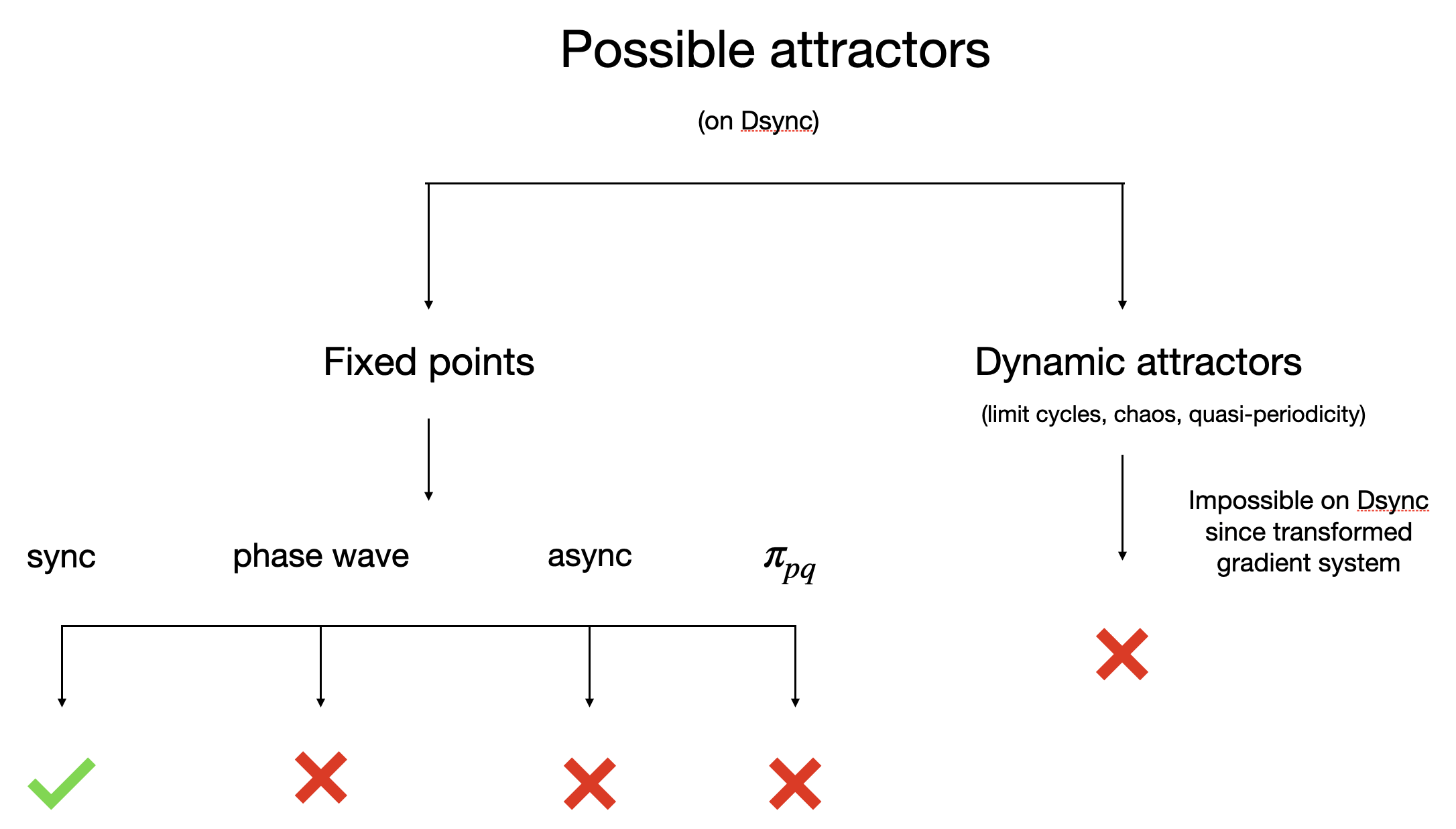}
\caption{Sketch of proof structure. $D_{sync}$ is the parameter region in which sync is known to be locally stable, and where we endeavour to show it is globally stable.}
\label{sketch}
\end{figure}

Figure~\ref{sketch} gives an outline of our proof's structure. It is a processes of elimination. We enumerate all the possible attractors that exist on $D_{sync}$, the parameter region where sync is known to be locally stable, and then show all but sync are unstable on this $D_{sync}$. 

Dynamic attractors (limit cycles, quasi-periodicity, chaos etc) are ruled out by showing the model has a transformed gradient structure on $D_{sync}$. Importantly, it does \textit{not} have this structure outside of $D_{sync}$. Transformed gradients are different to regular gradients, and to our knowledge are not widely talked about in nonlinear dynamics.

Fixed point attractors are the only remaining options. We \textit{prove} that aside from sync, there are only two possible categories of fixed points: those which lie on the async and phase wave manifolds, and what we call $\pi_{pq}$ points in which $(p,q)$ swarmalators have either $(\xi,\eta) = (C_1, C_2)$, and the remaining $(N-p,N-q)$ are $\pi$ units apart $(\xi,\eta) = (C_1+\pi, C_2 + \pi)$ (more later). We show the $\pi_{pq}$ points are unstable for all parameter values and are thus irrelevant to the steady state dynamics, leaving the fixed points on the phase wave and async manifolds as the only ones of interest. We show these are locally unstable on $D_{sync}$, leaving sync as the global attractor on $D_{sync}$ as claimed.

To sum up, the novelties of our analysis are the use of a transformed gradient structure, and the calculation of the stability spectra for an entirely manifold's worth of fixed points. This latter part has a microscale/macroscale structure that reminds us of statistical mechanics, and makes us think the $\lambda$'s may have use beyond our global sync theorem.

\section{Preliminaries}

\subsection{Transformed gradient system}
Recall a dynamical system $\dot{z} = f(z)$ where $z \in \mathbb{R}^n$ is a gradient system if it can be written
\begin{align}
    \dot{z} = - \nabla V(z)
\end{align}
for some gradient $V(z)$. A \textit{transformed gradient system} is a dynamical system which is not gradient by itself, but becomes one after a smooth coordinate transformation $\bar{z} = h(z)$.  It can be shown these have form \cite{pedergnana2022exact}
\begin{align}
    \dot{\bar{z}} = g^{-1}(z) \nabla V(\bar{z})
\end{align}
where $g^{-1}(z)$ is the inverse metric associated with the transformation $h(z)$. More formally, if $\mathcal{J} = \nabla h(z)$ be the Jacobian of the transformation $h(z)$, and $\mathcal{J} = Q b$ is the decomposition such that $Q = Q^T$, then $g = b b^T$ is a symmetric, positive definite metric. The restriction of $g$ being positive definite will be important.

The 1d swarmalator model is \textit{not} a gradient system, but as we will show it is a transformed gradient system. Although we mention in it does have a nice gradient/hamiltonian decomposition. Setting $z = (x_1, \dots, x_n, \theta_1, \dots \theta_n)$, we see
\begin{align}
    \dot{z} = - \nabla V + S \nabla H
\end{align}
where $S = ((0,1),(-1,0))$. The gradient is
\begin{align}
    V &= -\frac{K}{2N} \sum_{ij} \cos(\xi_j - \xi_i) + \cos(\eta_j - \eta_i) \\
      V                    &= -\frac{K}{2N} Re \sum_{ij} e^{i(\xi_j -\xi_i)} + e^{i(\eta_j -\eta_i)} \\  
      V                    &= -\frac{K}{2N} Re ( N Z_{\xi} N Z_{\xi}^*   + N Z_{\eta} N Z_{\eta}^*) \\  
    V                    &= -\frac{KN}{2}( r^2 + s^2) 
\end{align}
and the Hamiltonian is
\begin{align}
    H &= \frac{J}{2N} \sum_{ij} \cos(\xi_j - \xi_i) - \cos(\eta_j - \eta_i) \\
                          &=  \frac{JN}{2} (r^2 - s^2)
\end{align}
This is the Helmholtz decomposition for a `planar' system \cite{pedergnana2023certain} (planar in the sense $H(\xi,\eta)$ is a mean field quantity and depends on just two coordinates $(\xi,\eta)$). 

But this doesn't concern us here. 

\subsection{Routh-Hurwitz stability conditions}
Consider a dynamical system $\dot{z} = f(z)$ with a fixed point $z_0$ and characteristic equation
\begin{align}
    a_0 + a_1 \lambda + \dots + a_{N-1} \lambda^{N-1} + \lambda^N = 0 \label{char}
\end{align}
at that fixed point. The standard way to determine the stability of $z_0$ is to solve Eq.~\eqref{char} for $\lambda$ and check if all the real parts are negative. The Routh-Hurwitz conditions, by contrast, provide a way to check for stability without explicitly solving for $\lambda$. Instead, they give a set of conditions on the coefficients $a_i$. These conditions guarantee that all the roots of the characteristic equation have negative real parts, which corresponds to the stability of the fixed point $z_0$. 

For small values of $N$, the conditions are simple. 

For $N = 2$, the characteristic equation is $a_0 + a_1 \lambda + \lambda^2 = 0$. The Routh-Hurwitz stability conditions are:
\begin{align}
    a_1 > 0, \quad a_0 > 0
\end{align}
For $N = 3$, the characteristic equation is $a_0 + a_1 \lambda + a_2 \lambda^2 + \lambda^3 = 0$. The Routh-Hurwitz stability conditions are:
\begin{align}
    a_1 > 0, \quad a_2 > 0, \quad a_1 a_2 > a_0
\end{align}
For $N = 4$, the characteristic equation is $a_0 + a_1 \lambda + a_2 \lambda^2 + a_3 \lambda^3 + \lambda^4 = 0$. The Routh-Hurwitz stability conditions are:
\begin{align}
    a_1 > 0, \quad a_3 > 0, \quad a_2 a_3 > a_1, \quad a_1 a_2 a_3 > a_0 a_3^2
\end{align}

As $N$ increases the number of conditions increases. One key result that hold for any $N$ is that $a_{N-1} > 0$, the second to largest coefficient has to be positive. We will use this result to rule \textit{out} stability of various fixed point in the 1d swarmalator model.

Next we prove our main theorem.

\section{Results}

\subsection{Main theorem}

\begin{theorem}[Global Stability Theorem]
The sync manifold defined by
\begin{align}
    & r =  \Big| \frac{1}{N} \sum_j e^{ i \xi_j } \Big| = 1 \Longleftrightarrow \xi_i = C_1 \\
    & s =  \Big| \frac{1}{N} \sum_j e^{ i \eta_j } \Big| = 1 \Longleftrightarrow \eta_i = C_2
\end{align}
for $N>2$ is globally stable in the region $D_{sync} := ( J', K' |  J', K' > 0)$ or in transformed coordinates $(J,K | J-K < 0, J+K < 0 )$. Equivalently, for all random initial conditions except for a set of measure $0$, all trajectories converge to the sync fixed points defined by $\xi_i, \eta_i = C_1, C_2$ where the $C_1,C_2$ are constant determined by the initial conditions. 
\end{theorem}

Notice by "sync" we mean not just that all phases are identical, but all positions too. Note also that in $(x_i, \theta_i)$ coordinates the fixed point are $x_i \in [0, \pi]$ and $\theta_i \in [0, \pi]$ as previously discussed. The case $N=2$ is non-generic so we exclude it. Unless otherwise stated, assume $N>2$ throughout our analysis.

Next prove a series of lemmas to prove this theorem.

\begin{lemma}[Transformed gradient systems]
The 1d swarmalator model is a transformed gradient system in the sync region $D_{sync}$.  
\end{lemma}

\begin{proof}
It's easier to see this in original coordinates
\begin{align}
\dot{x}_i =  \frac{J'}{N} \sum_j \sin(x_j - x_i) \cos(\theta_j - \theta_i) \label{a1} \\
\dot{\theta}_i =  \frac{K'}{N} \sum_j \sin(\theta_j - \theta_i) \cos(x_j - x_i) \label{a2}
\end{align}
Defining the transformation as $X = x/\tilde{J}, T =  \theta / \tilde{K}$, then the metric is $g = (J', \dots J', K, \dots, K')$ and the gradient is 
\begin{align}
& V(Z) = -\frac{1}{2N} \sum_{ij} \cos (J' (X_j - X_i)) \cos (K' (T_j - T_i))
\end{align}
where $Z = (X_1, \dots X_N, T_1, \dots T_N)$. Then the 1d swarmalator model given by Eq~\eqref{a1},~\eqref{a2} is equivalent to
\begin{align}
\dot{\bar{z}} = g^{-1} \nabla \tilde{V}(\bar{z})
\end{align}
The key point here is that we require the metric to be positive definite which only holds when $J', K' > 0$ which is the sync region $D_{sync}$.
\end{proof}

The implication is that only fixed point attractors exist on $D_{sync}$. Limit cycles, chaos, and other non-stationary attractors are ruled out.

\begin{lemma}[]
For $J,K \neq 0$, all fixed point solutions of the 1d swarmalator model are either $\pi_{pq}$ states or lie on the async, phase wave, or sync manifolds defined by $(r,s) = (0,0), (1,0)/(0,1), (1,1)$.
\end{lemma}

\begin{proof}
Look at the model again where we have set $\phi = \psi = 0$ without loss of generality
\begin{align}
\dot{\xi}_i =  - K r \sin \xi - J s \sin \eta \label{y1} \\
\dot{\eta}_i =  - J r \sin \xi - K s \sin \eta \label{y2} 
\end{align}
If the swarmalator sit at fixed points, then $(r,s)$ must take stationary values. There are four options
\begin{itemize}
    \item (i) $(r,s) = (0,0)$
    \item (ii) $(r,s) = (a,0) \hspace{0.2cm} \text{or} \hspace{0.2cm} (0,a)$
    \item (iii) $(r,s) = (a,a)$
    \item (iv) $(r,s) = (a,b) \hspace{0.2cm} \text{or} \hspace{0.2cm} (b,a)$
\end{itemize}
Here $ 0 \leq a, b \leq 1$ and $a\neq b$. We must show each case collapses to either to $\pi_{pq}$ states, or the sync, phase wave or async manifold.

Case (i) describe the async manifold by definition. 

Case (ii) describes the phase wave manifold ($a=1$) or $\pi_{p0}$ states $0 \leq a \leq 1$ To see this, consider the governing Eqs.~\eqref{y1},~\eqref{y2} with $(a,0)$:
\begin{align}
\dot{\xi}_i =  - K a \sin \xi = 0  \\
\dot{\eta}_i =  - J a \sin \xi = 0
\end{align}
Since we assume $a>0$, the only solutions are $\sin \xi_i = 0 \Rightarrow \xi_i = 0, \pi$. If each swarmalator $i$ has $\xi_i = 0$ or $\xi_i = \pi$ for all $i$, then we are on the phase wave manifold, because constant $\xi$ implies $r = a = |\langle e^{i \xi} \rangle| \Rightarrow 1$.

The remaining cases where $p$ swarmalators have $\xi_i = 0$ and the remaining $N-p$ have $\xi_i = \pi$, defines the $\pi_{p0}$ state where $0 \leq a \leq 1$. Note here $0 < p < N$ and that the case $p=N/2$ for $N$ even is excluded; if $p=N/2$, then the half at $\xi=0$ and $\xi = \pi$ are equal and cancel out giving $a=0$ (the async manifold), but we assume $a>0$. So case (ii) describes the phase wave manifold or $\pi_{p0}$ states as claimed.

Case (iii) describes the sync manifold or the $\pi_{pq}$ states for $q\neq0$. Plugging $(r,s) = (a,a)$ into the equations we get
\begin{align}
\dot{\xi}_i =  - a( K \sin \xi + J \sin \eta ) = 0 \\
\dot{\eta}_i =  -a ( J \sin \xi + K  \sin \eta ) = 0
\end{align}
One solution here is $\sin \xi$ and $\sin \eta$ are zero simultaneously, which implies the sync state or the $\pi_{pq}$ states for the same reasons as case (ii). A second solution is one-dimensional manifold of fixed points. But this latter case necessitates the nullclines $\dot{\xi} = 0, \dot{\eta_i} = 0$ to describe the \textit{same} manifold. This is only possible in the special case $K=J$, or in original units $\tilde{J}+\tilde{K} = \tilde{J} - \tilde{K} \Rightarrow \tilde{K} = 0$. This is a marginal parameter setting we do not consider (first line of the lemma).

Finally, case (iv) also describes the sync manifold or the $\pi_{pq}$ states. Plugging $(r,s) = (a,b)$ into the equations yields
\begin{align}
\dot{\xi}_i =  - K a \sin \xi - J b \sin \eta = 0  \\
\dot{\eta}_i =  - J a \sin \xi - K b \sin \eta = 0
\end{align}
The situation is the same as above. Either the solutions as zero-dimensional fixed points which implies either $\pi_{pq}$ points, or the sync state $(a,b) = (1,1)$ which violates our assumption $a \neq b$, or there is a 1d manifold of fixed points $\Gamma(\xi, \eta) = 0$. But as before, this requires the nullclines describe the same curve, which is only possible in the special case $J=K \Rightarrow \tilde{K} = 0$ which we do not consider. 

Thus, the only fixed point solutions to the 1d swarmalator model for $J,K \neq 0$ are either $\pi_{pq}$ states, or those that lie on the async, phase wave, or sync manifolds as claimed
\end{proof}

Next we determine the parameter regions where each of these states is locally stable.

\begin{lemma}[Sync manifold stability]
The fixed points lying on the sync manifold defined by 
\begin{align}
    & r =  \Big| \frac{1}{N} \sum_j e^{ i \xi_j } \Big| = 1 \Longleftrightarrow \xi_i = C_1 \\
    & s =  \Big| \frac{1}{N} \sum_j e^{ i \eta_j } \Big| = 1 \Longleftrightarrow \eta_i = C_2
\end{align}
with $N>2$ has three distinct eigenvalues
\begin{align}
   & \lambda_0 = 0, \quad \text{wm } 1 \\
   & \lambda_1 = J+K, \quad \text{wm } N-1 \\
   & \lambda_2 = J-K, \quad \text{wm } N-1
\end{align}
which imply fixed points are locally stable only on $D_{sync}$.
\end{lemma}

\begin{proof}
The Jacobian at this fixed point has block form
\begin{equation}
    M_{sync} = \left[ 
\begin{array}{cc} 
  K A_0 & J A_0 \\ 
  J A_0 & K A_0 \\
\end{array} 
\right] \label{Mss}
\end{equation}
where
\begin{equation}
A_0 := \begin{bmatrix}
    - \frac{N-1}{N}   & \frac{1}{N}  & \dots & \frac{1}{N}\\
    \frac{1}{N}  & - \frac{N-1}{N} &  \dots & \frac{1}{N} \\
    \hdotsfor{4} \\
    \frac{1}{N}  & \frac{1}{N}  & \dots & - \frac{N-1}{N} 
\end{bmatrix} \label{A0a}
\end{equation}
Notice that the Jacobian of the entire system $M$ has $\dim = 2N$ since there are two state variables $(\xi, \eta)$ for each of the $N$ swarmalators, but that $\dim(A_0) = N$ since it is a sub-block of $M$. Now, the eigenvalues of $A_0$ are well known: there is $1$ with value $\hat{\lambda} = 0$ stemming from the rotational symmetry of the model and $N-1$ eigenvalues with $\hat{\lambda} = -1$. Recall the following identity for block matrices
\begin{equation}
    \det E := \det \left[ 
\begin{array}{cc} 
  C & D \\ 
  D & C \\
\end{array} 
\right] = \det(C+D) \det(C-D)
\end{equation}
This implies the eigenvalues of $E$ are the union of the eigenvalues of $C + D$ and $C-D$. Applying this identity to $M_{sync}$ (which has the required symmetric structure) yields
\begin{align}
    \lambda_0 &= 0 \\
    \lambda_1 &= - (K+J) \\
    \lambda_2 &= -(K-J)
\end{align}
with multiplicities $2, -1 + N, -1 + N$. The two zero eigenvalues corresponds to the rotational symmetry in the model and thus are \textit{fully} neutrally stable (zero at all orders, not just linear order) and not just \textit{linearly} neutrally stable. \footnote{By rotational symmetry we mean you can add a constant to each phase $\xi_i, \eta_i \rightarrow \theta_i + C_a, C_b$ for constants $C_a, C_b$ and the dynamics don't change (because only phase differences $\xi_j - \xi_i$ etc appear in the governing ODEs)}

Thus, sync is locally stable on $D_{sync}$ defined by $J+K>0, K-J>0$ as claimed. 
\end{proof}

\begin{lemma}[Stability of $\pi_{pq}$ states]
The $\pi_{pq}$ states are those were $p$ swarmaltors have $\xi=C_1$, the remaining $N-p$ have $\xi=C_1+\pi$, and $q$ swarmalators have $\eta = C_2$ and the remaining $N-q$ have $\eta = C_2 + \pi$. The eigenvalues of the states for $0 < p,q < N$ with $p,q \neq N/2$ for $N$ even are
\begin{align}
   \lambda_0 &= 0 \quad \text{wm } 2 \\
   \lambda_1 &= \frac{K+J}{2} \quad \text{wm } 1  \\
   \lambda_2 &= b_{pq} (K+J) \quad \text{wm } \frac{N}{2}-1 \\
   \lambda_3 &= -b_{pq} (K+J) \quad \text{wm } \frac{N}{2}-1\\
   \lambda_4 &= \frac{K-J}{2} \quad \text{wm } 1 \\
   \lambda_5 &= c_{pq} (K-J) \quad \text{wm } \frac{N}{2}-1 \\
   \lambda_6 &= -c_{pq} (K-J) \quad \text{wm } \frac{N}{2}-1
\end{align}
where $b_{pq}, c_{pq} > 0$ are constants that depend on $0<p,q<N$ (and $N>1$). These eigenvalues imply that the $\pi_{pq}$ states are unstable for all $J,K \neq 0$, since $\text{Re}(\lambda_2), \text{Re}(\lambda_4)$ have opposite signs to $\text{Re}(\lambda_3), \text{Re}(\lambda_5)$). Thus, $\pi_{pq}$ states are unstable on the sync region $D_{sync}$.
\end{lemma}

\begin{proof}
Straightforward algebra shows the characteristic equation takes the desired form:
\begin{align}
    & \lambda^2 
     \left(\lambda - \frac{K+J}{2}\right) 
     \left(\lambda + \frac{K+J}{2}\right)^{\frac{N}{2}-1} \times \\
    & \left(\lambda - b_{pq} (K+J)\right)^{\frac{N}{2}-1}  
     \left(\lambda + b_{pq} (K+J)\right)^{\frac{N}{2}-1} \times \\
    & \left(\lambda - c_{pq} (K-J)\right)^{\frac{N}{2}-1} 
     \left(\lambda + c_{pq} (K-J)\right)^{\frac{N}{2}-1} = 0
\end{align}
which proves the lemma. (The values $b_{pq}, c_{pq}$ take are unimportant).
\end{proof}

\begin{lemma}[Phase wave manifold stability]
All fixed points lying on the phase wave manifold defined by 
\begin{align}
    & r =  \Big| \frac{1}{N} \sum_j e^{ i \xi_j } \Big| = 0  \\
    & s =  \Big| \frac{1}{N} \sum_j e^{ i \eta_j } \Big| = 1 \Longleftrightarrow \eta_i = C_1
\end{align}
with $N>2$ have six distinct eigenvalues of form
\begin{align}
   & \lambda_0 = 0, \quad \text{wm } N-1 \\
   & \lambda_1 = -K, \quad \text{wm } N-3 \\
   & \lambda_2 = \frac{1}{4} \Big( -K (1+r_2) \pm \sqrt{ (r_2-3)^2 K^2  + 8 J^2(r_2-1)} \Big), \quad \text{wm} \hspace{0.1 cm} 1 \label{q1} \\
   & \lambda_3 = \frac{1}{4} \Big( -K (1-r_2) \pm \sqrt{ (r_2+3)^2 K^2  - 8 J^2(r_2+1)} \Big), \quad \text{wm} \hspace{0.1 cm} 1 \label{q2}
\end{align}
where $r_2 = \langle e^{2 i \xi} \rangle$. These are stable only on
\begin{align}
D_{wave} = \{(J, K) \ | \ K > 0 \ \text{and} \ J + K > 0 \ \text{and} \ J - K < 0 \}
\end{align}
and by extension unstable on $D_{sync}$. The sister manifold $(r,s) = (1,0)$ has the same stability region. Its eigenvalues are the same with $r_2 \rightarrow s_2 := \langle e^{2i \eta} \rangle$.
\end{lemma}

\textit{Remark}. The $N-1$ zero eigenvalues are fully, and not just linearly, neutral. This is because the dimension of the phase manifold is $\dim M_{wave} = N-1$ (Eq.~\eqref{dim_wave}). Movements in any of these $N-1$ dimensions does not change $r$ or $s$ which is what the zero eigenvalues signify. 

\begin{proof}
We study the $(r,s) = (0,1)$ branch of the phase wave manifold defined in the lemma (the analysis for $(r,s) = (1,0)$ branch is the same). The Jacobian here has block form:
\begin{align}
M_{wave} &= \begin{bmatrix}
K & A_1 & J & A_0 \\
J & A_1 & K & A_0
\end{bmatrix}
\end{align}
where $A_0$ is as before ~\eqref{A0a} and the new $A_1$ is
\begin{align}
    (A_1)_{ii} &= -\frac{1}{N} \sum_{j \neq i} \cos (\xi_j -\xi_i) \label{A1ii}\\
    (A_1)_{ij} &= \frac{1}{N} \cos (\xi_j -\xi_i)
\end{align}
We can make some simplifications. First, we assume without loss of generality that $\langle \xi \rangle = \langle \eta \rangle = 0$ (which corresponds to choice of frame). This implies $Z_{\xi}, Z_{\eta} = \langle e^{i \xi} \rangle, \langle e^{i \eta} \rangle \rightarrow r,s = \langle \cos \xi \rangle, \langle \cos \eta \rangle$. The benefit of all this is that the sums of cosines above can be expressed in terms of $r$. First notice the diagonal can be made simpler
\begin{align}
(A_1)_{ii} &= -\frac{1}{N} \sum_{j \neq i} \cos (\xi_j -\xi_i) \\
          &= -\frac{1}{N} \Big( -1 +   \sum_{j} \cos (\xi_j -\xi_i) \Big) \\
          &= -\frac{1}{N} \Big( -1 +   \sum_{j} \cos (\xi_j -\xi_i) \Big) \\
          &= -\frac{1}{N} \Big( -1 +  Re  \sum_{j} e^{i \xi_j} e^{- i \xi_i} \Big) \\
          &= -\frac{1}{N} \Big( -1 +   N r e^{- i \xi_i} \Big) \\
          &= -\frac{1}{N} \Big( -1 +  0 \Big) \\ 
          &= \frac{1}{N} 
\end{align}
where in the second to last line we used the fact that $r = 0$ on the phase wave manifold. Now comes the hard part. We want to find the characteristic equation $M_{wave}$ to find its eigenvalues $\lambda$. After much algebra, we find
\begin{align}
    \lambda^{N-1} (\lambda+K)^{N-3} (\lambda^4 + K \lambda^3 + a_2 \lambda^2 + a_1 \lambda + a_0 ) \label{t1}
\end{align}
We see the $N-1$ zero eigenvalues and the $N-3$ eigenvalues with $\lambda = -K$ claimed in the lemma pop out. What remains is thus to simplify the quartic expression. The coefficients $a_i$ are messy expressions that are unenlightening to display here (we provide links to our Mathematica notebooks if you want to view and interact with them \cite{github}). Their key features are that they depend on sums over cosines like the following
\begin{align}
    & S_1 = \sum_{j>i} \cos(\xi_j - \xi_i) \\
    & S_2 = \sum_{j>i} \cos(2(\xi_j - \xi_i)) \\
    & S_3 = \sum_{i=1}^{n} \sum_{j=i+1}^{n} \sum_{\substack{k=1 \\ k \neq i \\ k \neq j}}^{n} \cos(\xi_i + \xi_j - 2\xi_k)
\end{align}
 These can be simplified. Starting with $S_1$,
\begin{align}
S_1 &= \sum_{j>i} \cos(\xi_j - \xi_i)  = - \frac{N}{2} + \frac{1}{2} \sum_{i=1,j=1} \cos(\xi_j-\xi_i) \\
    &= -\frac{N}{2} + \frac{N^2}{2} \Big( \frac{1}{N^2} \sum_{i=1,j=1} \cos \xi_j \cos \xi_i + \sin \xi_j \sin \xi_i \Big) \\
    &= -\frac{N}{2}+ \frac{N^2}{2} r^2 \\
    &= -\frac{N}{2}
\end{align}
where on the last line we have used the fact that $r = 0$ on the phase wave manifold again. Similar calculations simplify $S_2, S_3$. Collecting all the results at once
\begin{align}
    S_1 &= -\frac{N}{2} \\
    S_2 &= - \frac{N}{2} + \frac{N^2}{2} r_2^2 \\
    S_3 &= N - \frac{N^2}{2} r_2^2 
\end{align}
where we remind you $r_2 = \langle \cos 2 \xi \rangle$. Plugging these expressions back into the quartic for $\lambda^4$ and simplifying we get
\begin{align}
& \lambda^4 + \lambda^3 K + \lambda^2 \left(J^2 - \frac{1}{4} K^2 \left(r_2^2 + 3\right)\right) \\
& + \frac{1}{2} \lambda K \left(r_2^2 + 1\right) \left(J^2 - K^2\right) - \frac{1}{4} \left(r_2^2 - 1\right) \left(J^2 - K^2\right)^2 = 0
\end{align}
This quartic has clean solution
\begin{align}
&\lambda_2 = \frac{1}{4} \Big( -K (1+r_2) \pm \sqrt{ (r_2-3)^2 K^2  + 8 J^2(r_2-1)} \Big)  \\
& \lambda_4 = \frac{1}{4} \Big( -K (1-r_2) \pm \sqrt{ (r_2+3)^2 K^2  - 8 J^2(r_2+1)} \Big)
\end{align}
as claimed which completes our proof. \end{proof}

\begin{figure}[t!]
\centering
\includegraphics[width = 0.8\columnwidth]{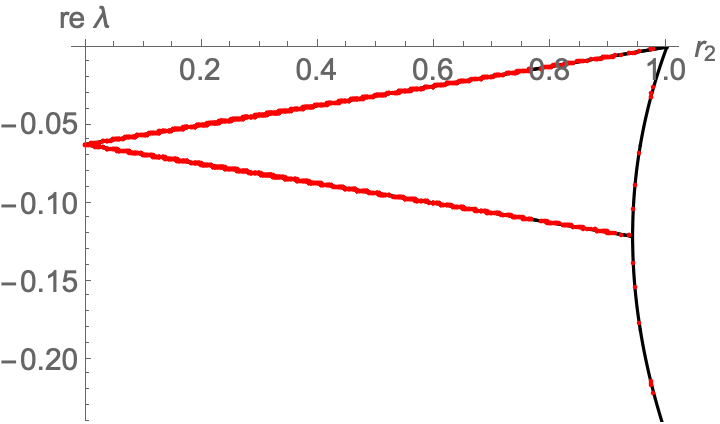}
\caption{Comparison of theoretical eigenvalues (black curve)  to empirically derived $\lambda$ (red dots). Notice there are red dots on each black curve, including the rightmost branch. To find the empirical eigenvalues, we ran simulations with $n=6$ swarmalators with $(K,J) = ()$ and fed the fixed points $\xi_i, \eta_i$ into the Jacobian and computed the eigenvalues numerically. We did this $N_{trial} = 10^3$ times. Note we just plot the two sets of complex conjugates $\lambda$'s given by Eq.~\eqref{q1},~\eqref{q2}; the $\lambda = -K,0$ are not plotted.}
\label{lam-pw}
\end{figure}

\textit{Remarks}. Figure~\ref{lam-pw} confirms our analytic prediction numerically by plotting $re(\lambda)$ versus $r_2$. See the caption for details on how the $\lambda$ were found numerically.

Recall that the $\lambda$'s are valid for \textit{all} microstates on the phase wave manifold -- they represent a family's worth of fixed points, not just one fixed point. The magnitude of the $\lambda$ for each of these fixed point depends on the value of $r_2$ value at that fixed point (which varies as you move around the phase wave manifold). Yet the stability properties of each fixed point are the same. They are all locally stable on
\begin{align}
D_{wave} := (J,K| K>0, J+K > 0, J-K < 0)   
\end{align}
as indicated by Figure~\ref{states-identical}(d). Finally, notice in the special case of uniform spacing $\xi_i = 2 \pi i / N$, which implies $r_2=0$, the expressions simplify to
\begin{align}
    & \lambda_0 = 0, \text{wm } N-1  \\
    & \lambda_1 = -K, \text{wm } N-3  \\
    & \lambda_2 = \frac{1}{4} \Big( -K \pm \sqrt{ 9 K^2 - 8 J^2} \Big), \text{wm } 2
\end{align}
which were reported previously \cite{o2022collective}. 

\begin{lemma}[Stability of $\pi_{p0}, \pi_{0p}$ states]
The $\pi_{p0}$ states are those were $p$ swarmalators have $\xi_i=C$ and the remaining $N-p$ have $\xi = C+\pi$. Their $\eta_i$ are fully desynchronized $s = 0$. The $\pi_{0p}$ states are the same with the roles of $\xi,\eta$ swapped. Both states are unstable on $D_{sync}$
\end{lemma}

\begin{proof}
Numerics indicate that these states are unstable for all the parameter values. Proving this was tricky because the characteristic equations for the eigenvalues $\lambda$ had both an $N$ and $p$ dependence. However after much experimentation, patterns emerged. To cut to the chase, we found two classes of characteristic equations separated by a boundary $p^*(N)$
\begin{align}
& \lambda^{N-1} \Big( a_0 + a_1 \lambda + \dots - \mu(N,p) K \lambda^{N} + \lambda^{N+1} \Big) = 0 ,  \text{for } p \leq p^* \label{f1} \\
& \lambda^{N-1} \Big( \lambda + C(N,p) K \Big)^{N-m} \Big( b_0 + b_1 \lambda + \lambda^2 \Big) \times \\ & \hspace{0.85 cm} \Big( c_0 + c_1 \lambda + c_2 \lambda^2 - \nu(N,p) K \lambda^{m-1} + \lambda^m \Big) = 0,  \text{for} p > p^* \label{f2}
\end{align}
with 
\begin{align}
    p^*(N) &= \frac{1}{2}(N - \sqrt{2 N}) \\ 
    \mu(N,p) &=  N - 2 - 4p + \frac{4p^2}{N} \label{thresh} \\
    C(N,p) &= \frac{1}{2} \Big(1 - \frac{2p}{N} \Big) \\
    \nu(N,p) &= \mu(N,p) - (N-m) C(N,p) 
\end{align}
where $m$ is an exponent that depends on $p$: $(p,m) = (1,4), (2,5), (3,6)$ and for $p>3$ $m=6$. 

In Eq.~\ref{f1}, the coefficient $a_i=a_i(J,K, \xi_1, \dots \xi_2)$ are complex expressions whose form is unimportant. All that matters is the $\mu(N,p)$ term which is strictly positive $\mu(N,p) > 0$ for $p \leq p*$ (we will derive it shortly). Recall a necessary condition for stability (from the Routh Hurwitz conditions) is the $n-1$-th term is positive $c_{n-1} > 0$. For our problem, since $\mu(N,p) >0$ this requires $K<0$. Thus, the $\pi_{p0}, \pi_{0p}$ states are \textit{unstable} on $K>0$ which constrains the sync region $D_{sync}$. (We believe they are unstable for all $K$, but we don't need to prove that here). 

The case $p > p^*(N)$, is the same. You can see the $n-1$-th term in the third polynomial is positive only when $K<0$ proving instability on $K>0$ and in turn $D_{sync}$. Taken together, these show the $\pi_{p0}$ states are unstable on $D_{sync}$ as claimed.

Now we show how we derived the characteristics equations Eqs.~\eqref{f1}, ~\eqref{f2}. The Jacobian of the $\pi_{p0}$ states is similar to $M_{wave}$ and has form
\begin{align}
M_{\pi} &= \begin{bmatrix}
K & A_1 & J & A_2 \\
J & A_1 & K & A_2
\end{bmatrix}
\end{align}
where $A_1$ is as before Eq.~\eqref{A1ii}. The new $A_2$ is similar to the $A_0$ Eq.~\eqref{A0a} but with some of the elements swapped
\begin{align}
(A_2)_{ii} &= -\frac{1}{N} \sum_{j \neq i} \cos(\xi_j - \xi_i) = \begin{cases}
-\frac{p-1}{N} + \frac{N-p}{N} & \text{if } \xi_i = 0, \\
-\frac{N-p-1}{N} + \frac{p}{N} & \text{if } \xi_i = \pi,
\end{cases} \\
(A_2)_{ij} &= - \cos(\xi_j - \xi_i) = \begin{cases}
-1 & \text{if } \xi_j = \xi_i, \\
1 & \text{if } \xi_j \neq \xi_i.
\end{cases}
\end{align}
Now we computing $\det(M_{\pi} - \lambda I)$. Straightforward algebra shows the $\lambda^{N-1}$ pre factor pops out, leaving a polynomial of degree $N-1$ in $\lambda$. We need to show the second to last term is $ K \mu(N,p) \lambda^{N-1}$. Recall this second to last term is minus the trace of the matrix $M_{\pi}$
\begin{align}
    K \mu(N,p) = -Tr(M_{\pi})
\end{align}
Since $M_{\pi}$ is block diagonal $Tr(M) = K Tr(A_1) + K Tr(A_2)$. We find $Tr(A_1) = 1$ and $Tr(M_2) =N-3-4p + 4p^2/N $. Putting this together gives the desired
\begin{align}
    K \mu(N,p) = -K \left( N - 2 - 4p + \frac{4p^2}{N} \right) \\
    \mu(N,p) = -\left( N - 2 - 4p + \frac{4p^2}{N} \right)
\end{align}
To summarize, what we have derived is the characteristic equation is
\begin{align}
    a_0 + a_1 \lambda + \dots - K \mu(N,p) \lambda^{N-1} + \lambda^N = 0
\end{align}
This is true for all $p$, but you can see if $\mu(N,p) > 0$ then we can use the RH conditions to rule out stability on $K>0$ which contains the sync region. Solving $\mu(p,N) = 0$ gives us the threshold $p^*N$ at which $\mu(N,p)$ flips from positive to negative which is Eq.~\eqref{thresh}. This explains Eq.~\eqref{f1}.

For $p>p^*$, this equation is still valid, but the (necessary but not sufficient) RH condition for stability $a_{n-1}>0$ is \textit{met} for $K>0$ which contains $D_{sync}$. So we need to find another way to rule \textit{out} stability on $K>0$. We did this by finding the root $\lambda - C(N,p)K$ via trial and error for $p>p^*$. Then we factored this out giving the equation ~\eqref{f2}, which had its $n-1$-th term $\nu(N,p) > 0$ for $p>p^*$ as we required. 

That completes our proof of this lemma.
\end{proof}

\textit{Remark}. We conjecture that the $\pi_{p0}, \pi_{0p}$ states are unstable for \textit{all} parameters values; proving this is an open problem.

\begin{lemma}[Async manifold stability]
All fixed points lying on the async manifold defined by
\begin{align}
    & r =  \Big| \frac{1}{N} \sum_j e^{ i \xi_j } \Big| = 0 \\
    & s =  \Big| \frac{1}{N} \sum_j e^{ i \eta_j } \Big| = 0 
\end{align}
with $N>2$ have five distinct eigenvalues. These satisfy
\begin{align}
    \lambda^{2N-4}(\lambda^4 + K \lambda^3 + a_2 \lambda^2 + a_1 \lambda + a_0) 
\end{align}
where the coefficients $a_i$ are complex expressions involving sums of trig terms. The Routh-Hurwitz conditions imply these are unstable for $K>0$, and by extension unstable on $D_{sync}$.
\noindent
\end{lemma}

\begin{proof}
The same as before. We compute the Jacobian and evaluate it on async manifold. This has form
\begin{align}
M_{async} &= \begin{bmatrix}
K & A_1(\xi) & J & A_1(\eta) \\
J & A_1(\xi) & K & A_1(\eta)
\end{bmatrix}
\end{align}
where the $A_1$ is as before. The characteristic equation is
\begin{align}
    \lambda^{2N-4}(\lambda^4 - 2 K \lambda^3 + a_2 \lambda^2 + a_1 \lambda + a_0) 
\end{align}
We were unable to simplify the terms $a_2, a_1, a_0$ in terms of trig moments. But we don't actually \textit{need} to do that here. We just need to show these are unstable on $D_{sync}$. You can see the $n-1$'th term $a_3 = -2K > 0$ only when $K<0$ which rules out stability on the sync region which completes the proof.
\end{proof}

\begin{proof}[Proof of global synchronization theorem]
We now have all the pieces of the puzzle in place to prove our theorem. Since the swarmalator model has a transformed gradient structure on $D_{sync}$, it can only have fixed point attractors (lemma 1) The only attractors are sync, the $\pi_{pq}$ states, and the phase wave an async manifolds (lemma 2). Sync is locally stable on $D_{sync}$ (lemma 3), but the latter are all unstable on $D_{sync}$ (lemmas 4,5,6,7), leaving sync as the single, global attractor on $D_{sync}$. Thus, for all random initial conditions except for a set of measure $0$ sync is globally attracting on $D_{sync}$. (The set of measure zero corresponds to initial condition that lie exactly on the unstable fixed points).
\end{proof}

\textit{Remark}. Notice we have not proved the global stability of the phase wave and async manifolds. The system does not have a gradient structure in those parameter regions, so we cannot rule out the existence of non-stationary attractors.

\subsection{Global stability of Kuramoto model}
We can use the technique defined above to provide a new proof for the global stability of the sync and async states in the homogenous Kuramoto model
\begin{align}
    & \dot{\theta}_i = \frac{K}{N} \sum_j \sin(\theta_j - \theta_i) \\
    & \dot{\theta}_i = -  K r \sin \theta 
\end{align}
Here $r e^{i \phi} := \langle e^{i \theta} \rangle $ is the sync order parameters (we are abusing notation here, since we used $r$ as one of the rainbow order parameters in the above too). We assume $\phi = 0$ without loss of generality.

Numerics indicate there are only two states: sync with all oscillators identical phases corresponding to $r=1$, and async corresponding to $r=0$. The first proof of the global stability of both of these states were provided by Watanabe and Strogatz via their famous WS equations \cite{watanabe1993integrability}. Others exist \cite{kassabov2022global}. But to our knowledge, none of these proofs provide the full spectrum of the async manifold (Eq.~\eqref{km-async}) like ours does.

Since the proof follows the same structure as above, we don't use the formal lemma/theorem structure; we just verbally explain instead. The calculations of the eigenvalues are also in essence the same as above, so we just quote results.

\begin{figure}[t!]
\centering
\includegraphics[width = 0.8\columnwidth]{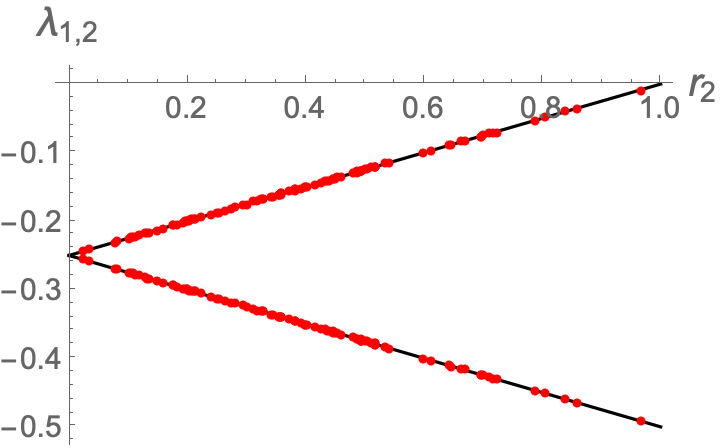}
\caption{Comparison of theoretical eigenvalues (black curve) to empirically derived $\lambda$ (red dots). Here $n=6$ oscillators with $k=-0.5$ were simulated $N_{trial} = 100$ times. The fixed points $\theta_i^*$ were collected, and fed into the Jacobian and the eigenvalues were computed numerically.}
\label{lam-km}
\end{figure}

The Kuramoto model is a gradient systems for all $K$ so its only attractors are fixed points. There are three classes of fixed points, those on the sync manifold with $r=1$, those on the async manifold $r=0$, and $\pi_{p}$ states. The eigenvalues of the sync state are
\begin{align}
\lambda &= -K \\
\lambda &= 0 
\end{align}
which imply $D_{sync} := \{K| K>0\}$. Those of the $\pi_p$ states are
\begin{align}
\lambda_1 &= 0 \\
\lambda_2 &= K \\
\lambda_3 &= - \frac{N-2p}{N} p
\end{align}
These are unstable for all $K$. Finally, those for the async manifold are\begin{align}
    &\lambda_0 = 0\\
    & \lambda_1 = \frac{K}{2} \Big( 1 \pm r_2 \Big) \label{km-async}
\end{align}
which are locally stable on $D_{async} := \{K|K<0\}$ (the $0 \leq r_2 \leq$ changes the magnitude of $\lambda_1$ but not the sign). As before  $r_2 := | \langle \exp(2i\theta) \rangle |$. Figure~\ref{lam-km} shows our analytic predictions for $\lambda$ match numerics.

To sum up, since the local stability regions of the async and sync manifolds are non-overlapping, and the $\pi_p$ state is always unstable, the local stability of each state implies global stability of each state, as claimed.

\section{Discussion}
We have proved the global stability of the sync state for perhaps the simplest possible model of swarmalators: mean field coupling with the motion confined to a 1d periodic domain. Because our work is intended as a starting point for a larger effort, we elaborate at length on the different directions that future work can take.

\subsection{Global stability of the async and phase wave manifolds.}
Two natural next questions are the global stability of the phase wave and the async manifolds. The mathematical difficulty here is our model loses its transformed gradient structure in the parameter regions where phase waves and async are locally stable $D_{wave}, D_{sync}$ so we cannot rule out the existence of non-stationary attractors. In fact, we know the phase wave destabilizes via Hopf bifurcation; numerics indicate this is sub-critical, meaning an unstable limit cycle co-exists with the phase wave state. So, one would need to come up with a Lyapunov function which would take some ingenuity.

Our hunch is the Helmholtz decomposition we identified earlier could be useful here. Recall this is
\begin{align}
    \dot{z} = - \nabla V + S \nabla H
\end{align}
where $S = ((1,0),(-1,0))$ and
\begin{align}
    V(\xi,\eta) &= -\frac{N K}{2} (r^2 + s^2) \\
    H(\xi,\eta) &= \frac{JN}{2} (r^2 - s^2)
\end{align}
It's clean form looks highly suggestive to us -- it ought to be useful somehow. Consider, for example, the manifold defined by $r=s$. Here, the Hamiltonian piece turns off $H = 0$ and so the motion is purely gradient (which rules out nonstationary orbits as we hoped), so we ought to be able to prove global stability there. Then perhaps by considering small perturbations  of this `gradient manifold`, we might be able to guess the form of a Lyapunov function for the entire phase space.

There is also the issue of finding the stability spectrum of the async manifold. Recall that we derived the characteristic equation $\lambda^{2N-4} (\lambda^4 - 2K \lambda^3 + a_2 \lambda^2 + a_1 \lambda + a_0) = 0$. We were unable to simplify the coefficient $a_2,a_1,a_0$ in terms of trig moments like $r_{2}, s_{2}, r_{11}, s_{11}$. The expressions $a_i$ were monsters -- long strings of terms like $\sin(\xi_i -\eta_j - 2 \eta_k)$; take a look at our Mathematica notebooks \cite{khev2024swarmalators}. Our guess is that only the four mentioned moments arise (and not the higher ones like $r_{3}$ or $s_{2,3}$) and that $a_i$ will contain triple products like $r_2 s_2 r_{11}$.

\subsection{Realism}
However, in order for any global stability results to be genuinely useful in the lab and beyond, we would need to add more realism to the model. Bettstetter and colleagues, who have actually implemented swarmalatorbots in the lab \cite{barcis2020sandsbots}, have identified three key features: delayed coupling, local coupling, and stochastic coupling \cite{schilcher2021swarmalators}. And, of course, we would need to consider motion in the plane or in open space $x_i \in \mathbb{R}^2, \mathbb{R}^3$.

\textbf{Delays} are easy to model. You can just add a delay $\tau$ to the 1d model, 
\begin{align}
\dot{x}_i &= \nu +   \frac{J'}{N} \sum_j \sin(x_j(t-\tau) - x_i) \cos(\theta_j(t-\tau) - \theta_i)  \\
\dot{\theta}_i &= \omega + \frac{K'}{N} \sum_j \sin(\theta_j(t-\tau) - \theta_i) \cos(x_j(t-\tau) - x_i),
\end{align}
and start analyzing (note we have left the identical frequencies in since you can't remove them via a change of frame anymore). For an ODE of form \( \dot{x}(t) = A x(t) + B x(t - \tau) \) with $x \in \mathbb{R}^N$, the eigenvalues satisfy the equation \( \det(\lambda I - A - B e^{-\lambda \tau}) = 0 \). Applying this to the sync state should be straightforward. The Kuramoto model with time delays has also already been analyzed \cite{yeung1999time}; their analysis would be helpful to review. After that, you can work up to the 2d model. Some results in this direction have been reported \cite{blum2024swarmalators}. 

\textbf{Local coupling} can be tackled by adding a box function of radius $\sigma'$ to the governing equations like so 
\begin{align}
\dot{x}_i = \frac{J'}{N} \sum_j \sin(x_j - x_i) \cos(\theta_j - \theta_i) G_{\sigma'}(x_j - x_i) \label{tt1} \\
\dot{\theta}_i = \frac{K'}{N} \sum_j \sin(\theta_j - \theta_i) \cos(x_j - x_i) G_{\sigma'}(x_j - x_i) \label{tt2}
\end{align}
For ease set $J'=K'=1$ (which puts us in the sync region). Then define $\sigma = \sigma'/\pi$, so that sweeping $\sigma$ from $1$ to $0$ corresponds to full range to zero range. Is there a lower threshold $\sigma_l$ below which sync is never realized, and an upper threshold $\sigma_u$ above which sync is guaranteed to occur, starting from random initial conditions? 

\textbf{Stochastic coupling} is harder to deal with. You need to discretize time, which takes us out of the familiar world of smooth coupling modelable with ODEs. Imagine at every timestep $\delta t$ \footnote{with $\delta t$ large enough that the $\delta t \rightarrow 0$ limit is inappropriate, and so we can't use ODE approximations} that each robot in a swarm fires a communication pulse $P$. Define the right hand side of the governing equations like $\dot{x}_i, \dot{\theta}_i = f(x,\theta), g(x,\theta)$. Then this pulse is $P = (f(x,\theta), g(x,\theta))$. Now, say each pulse $P$ fails with probability $r$. How does the liklilood of synchrony depend on $r \in [0,1]$? Schilcher and Bettstetter have studied this numerically \cite{schilcher2021swarmalators}, but how could one approach it analytically? 

Our instinct is to retreat a little, and instead of trying to analyze the swarmalators model directly (like Schilcher and Bettstetter did) consider the simpler Peskin model \cite{peskin1975mathematical}. This famous model considers a population of $N$ integrate-and-fire oscillators. Each oscillator climbs up a phase curve defined by $\dot{\theta}_i = S_0 - \gamma \theta_i$, and when it reached the threshold value $\theta = 1$ it fires a pulse of size $\epsilon$ which is instantly felt by all other oscillators. Then it resets its phase back to zero $\theta_i \rightarrow 0$. The simplicity of the model allows one to prove global synchrony will occur for all $\epsilon>0$ and $\gamma>0$, when the oscillators are identical and globally coupled \cite{mirollo1990synchronization}. Its transient dynamics can be described using aggregation theory \cite{o2015synchronization,o2016transient,gwon2023percolation}.

Now say the units move around in 1d according to some spatial dynamics. When a pulse is received, this alters their $x$'s in some way -- perhaps pushes them a distance $ \epsilon$ in the direction of whichever swarmalator sent the pulse. What is the probability that such Peskin swarmalators end up in sync in both phase $\theta_i = \theta_j$ and space $x_i = x_j$ given a pulse strength $\epsilon$? And how would this change under local coupling of range $\sigma \in [0,1]$ and drop-rate/stochastic coupling $r \in [0,1]$? Numerical experiments should be easy to run here, and might give some clues on how to attack the problem analytically.

Any results in this direction would be a big step forward in connecting the sync community to the applied scientists making use of sync-selected self-assembly -- our main goal, recall, as articulated in the \textit{Introduction}.

\bibliographystyle{apsrev}
\bibliography{ref}

\begin{thebibliography}{67}
\expandafter\ifx\csname natexlab\endcsname\relax\def\natexlab#1{#1}\fi
\expandafter\ifx\csname bibnamefont\endcsname\relax
  \def\bibnamefont#1{#1}\fi
\expandafter\ifx\csname bibfnamefont\endcsname\relax
  \def\bibfnamefont#1{#1}\fi
\expandafter\ifx\csname citenamefont\endcsname\relax
  \def\citenamefont#1{#1}\fi
\expandafter\ifx\csname url\endcsname\relax
  \def\url#1{\texttt{#1}}\fi
\expandafter\ifx\csname urlprefix\endcsname\relax\def\urlprefix{URL }\fi
\providecommand{\bibinfo}[2]{#2}
\providecommand{\eprint}[2][]{\url{#2}}

\bibitem[{\citenamefont{Tan et~al.}(2022)\citenamefont{Tan, Mietke, Li, Chen, Higinbotham, Foster, Gokhale, Dunkel, and Fakhri}}]{tan2022odd}
\bibinfo{author}{\bibfnamefont{T.~H.} \bibnamefont{Tan}}, \bibinfo{author}{\bibfnamefont{A.}~\bibnamefont{Mietke}}, \bibinfo{author}{\bibfnamefont{J.}~\bibnamefont{Li}}, \bibinfo{author}{\bibfnamefont{Y.}~\bibnamefont{Chen}}, \bibinfo{author}{\bibfnamefont{H.}~\bibnamefont{Higinbotham}}, \bibinfo{author}{\bibfnamefont{P.~J.} \bibnamefont{Foster}}, \bibinfo{author}{\bibfnamefont{S.}~\bibnamefont{Gokhale}}, \bibinfo{author}{\bibfnamefont{J.}~\bibnamefont{Dunkel}}, \bibnamefont{and} \bibinfo{author}{\bibfnamefont{N.}~\bibnamefont{Fakhri}}, \bibinfo{journal}{Nature} \textbf{\bibinfo{volume}{607}}, \bibinfo{pages}{287} (\bibinfo{year}{2022}).

\bibitem[{\citenamefont{Quillen et~al.}(2021)\citenamefont{Quillen, Peshkov, Wright, and McGaffigan}}]{quillen2021metachronal}
\bibinfo{author}{\bibfnamefont{A.}~\bibnamefont{Quillen}}, \bibinfo{author}{\bibfnamefont{A.}~\bibnamefont{Peshkov}}, \bibinfo{author}{\bibfnamefont{E.}~\bibnamefont{Wright}}, \bibnamefont{and} \bibinfo{author}{\bibfnamefont{S.}~\bibnamefont{McGaffigan}}, \bibinfo{journal}{Physical Review E} \textbf{\bibinfo{volume}{104}}, \bibinfo{pages}{014412} (\bibinfo{year}{2021}).

\bibitem[{\citenamefont{Hrabec et~al.}(2018)\citenamefont{Hrabec, K{\v{r}}i{\v{z}}{\'a}kov{\'a}, Pizzini, Sampaio, Thiaville, Rohart, and Vogel}}]{hrabec2018velocity}
\bibinfo{author}{\bibfnamefont{A.}~\bibnamefont{Hrabec}}, \bibinfo{author}{\bibfnamefont{V.}~\bibnamefont{K{\v{r}}i{\v{z}}{\'a}kov{\'a}}}, \bibinfo{author}{\bibfnamefont{S.}~\bibnamefont{Pizzini}}, \bibinfo{author}{\bibfnamefont{J.}~\bibnamefont{Sampaio}}, \bibinfo{author}{\bibfnamefont{A.}~\bibnamefont{Thiaville}}, \bibinfo{author}{\bibfnamefont{S.}~\bibnamefont{Rohart}}, \bibnamefont{and} \bibinfo{author}{\bibfnamefont{J.}~\bibnamefont{Vogel}}, \bibinfo{journal}{Physical Review Letters} \textbf{\bibinfo{volume}{120}}, \bibinfo{pages}{227204} (\bibinfo{year}{2018}).

\bibitem[{\citenamefont{Liu et~al.}(2021)\citenamefont{Liu, Shi, Zhao, Chat{\'e}, Shi, and Zhang}}]{liu2021activity}
\bibinfo{author}{\bibfnamefont{Z.~T.} \bibnamefont{Liu}}, \bibinfo{author}{\bibfnamefont{Y.}~\bibnamefont{Shi}}, \bibinfo{author}{\bibfnamefont{Y.}~\bibnamefont{Zhao}}, \bibinfo{author}{\bibfnamefont{H.}~\bibnamefont{Chat{\'e}}}, \bibinfo{author}{\bibfnamefont{X.-q.} \bibnamefont{Shi}}, \bibnamefont{and} \bibinfo{author}{\bibfnamefont{T.~H.} \bibnamefont{Zhang}}, \bibinfo{journal}{Proceedings of the National Academy of Sciences} \textbf{\bibinfo{volume}{118}} (\bibinfo{year}{2021}).

\bibitem[{\citenamefont{Tanaka}(2007)}]{tanaka2007general}
\bibinfo{author}{\bibfnamefont{D.}~\bibnamefont{Tanaka}}, \bibinfo{journal}{Physical Review Letters} \textbf{\bibinfo{volume}{99}}, \bibinfo{pages}{134103} (\bibinfo{year}{2007}).

\bibitem[{\citenamefont{Iwasa et~al.}(2012)\citenamefont{Iwasa, Iida, and Tanaka}}]{iwasa2012various}
\bibinfo{author}{\bibfnamefont{M.}~\bibnamefont{Iwasa}}, \bibinfo{author}{\bibfnamefont{K.}~\bibnamefont{Iida}}, \bibnamefont{and} \bibinfo{author}{\bibfnamefont{D.}~\bibnamefont{Tanaka}}, \bibinfo{journal}{Physics Letters A} \textbf{\bibinfo{volume}{376}}, \bibinfo{pages}{2117} (\bibinfo{year}{2012}).

\bibitem[{\citenamefont{O'Keeffe et~al.}(2017)\citenamefont{O'Keeffe, Hong, and Strogatz}}]{o2017oscillators}
\bibinfo{author}{\bibfnamefont{K.~P.} \bibnamefont{O'Keeffe}}, \bibinfo{author}{\bibfnamefont{H.}~\bibnamefont{Hong}}, \bibnamefont{and} \bibinfo{author}{\bibfnamefont{S.~H.} \bibnamefont{Strogatz}}, \bibinfo{journal}{Nature Communications} \textbf{\bibinfo{volume}{8}}, \bibinfo{pages}{1} (\bibinfo{year}{2017}).

\bibitem[{\citenamefont{Adorj{\'a}ni et~al.}(2024)\citenamefont{Adorj{\'a}ni, Lib{\'a}l, Reichhardt, and Reichhardt}}]{adorjani2024motility}
\bibinfo{author}{\bibfnamefont{B.}~\bibnamefont{Adorj{\'a}ni}}, \bibinfo{author}{\bibfnamefont{A.}~\bibnamefont{Lib{\'a}l}}, \bibinfo{author}{\bibfnamefont{C.}~\bibnamefont{Reichhardt}}, \bibnamefont{and} \bibinfo{author}{\bibfnamefont{C.}~\bibnamefont{Reichhardt}}, \bibinfo{journal}{Physical Review E} \textbf{\bibinfo{volume}{109}}, \bibinfo{pages}{024607} (\bibinfo{year}{2024}).

\bibitem[{\citenamefont{Kongni et~al.}(2023)\citenamefont{Kongni, Nguefoue, Njougouo, Louodop, Ferreira, Tchitnga, and Cerdeira}}]{kongni2023phase}
\bibinfo{author}{\bibfnamefont{S.~J.} \bibnamefont{Kongni}}, \bibinfo{author}{\bibfnamefont{V.}~\bibnamefont{Nguefoue}}, \bibinfo{author}{\bibfnamefont{T.}~\bibnamefont{Njougouo}}, \bibinfo{author}{\bibfnamefont{P.}~\bibnamefont{Louodop}}, \bibinfo{author}{\bibfnamefont{F.~F.} \bibnamefont{Ferreira}}, \bibinfo{author}{\bibfnamefont{R.}~\bibnamefont{Tchitnga}}, \bibnamefont{and} \bibinfo{author}{\bibfnamefont{H.~A.} \bibnamefont{Cerdeira}}, \bibinfo{journal}{Physical Review E} \textbf{\bibinfo{volume}{108}}, \bibinfo{pages}{034303} (\bibinfo{year}{2023}).

\bibitem[{\citenamefont{Degond et~al.}(2022)\citenamefont{Degond, Diez, and Walczak}}]{degond2022topological}
\bibinfo{author}{\bibfnamefont{P.}~\bibnamefont{Degond}}, \bibinfo{author}{\bibfnamefont{A.}~\bibnamefont{Diez}}, \bibnamefont{and} \bibinfo{author}{\bibfnamefont{A.}~\bibnamefont{Walczak}}, \bibinfo{journal}{Analysis and applications} \textbf{\bibinfo{volume}{20}}, \bibinfo{pages}{1215} (\bibinfo{year}{2022}).

\bibitem[{\citenamefont{Degond and Diez}(2023)}]{degond2023topological}
\bibinfo{author}{\bibfnamefont{P.}~\bibnamefont{Degond}} \bibnamefont{and} \bibinfo{author}{\bibfnamefont{A.}~\bibnamefont{Diez}}, \bibinfo{journal}{Acta Applicandae Mathematicae} \textbf{\bibinfo{volume}{188}}, \bibinfo{pages}{18} (\bibinfo{year}{2023}).

\bibitem[{\citenamefont{Gong et~al.}(2024)\citenamefont{Gong, Zhou, and Huang}}]{gong2024approximating}
\bibinfo{author}{\bibfnamefont{Z.}~\bibnamefont{Gong}}, \bibinfo{author}{\bibfnamefont{J.}~\bibnamefont{Zhou}}, \bibnamefont{and} \bibinfo{author}{\bibfnamefont{M.}~\bibnamefont{Huang}}, \bibinfo{journal}{International Journal of Bifurcation and Chaos} \textbf{\bibinfo{volume}{34}}, \bibinfo{pages}{2450129} (\bibinfo{year}{2024}).

\bibitem[{\citenamefont{Smith}(2024)}]{smith2024swarmalators}
\bibinfo{author}{\bibfnamefont{L.~D.} \bibnamefont{Smith}}, \bibinfo{journal}{SIAM Journal on Applied Dynamical Systems} \textbf{\bibinfo{volume}{23}}, \bibinfo{pages}{1133} (\bibinfo{year}{2024}).

\bibitem[{\citenamefont{Levis et~al.}(2019)\citenamefont{Levis, Pagonabarraga, and Liebchen}}]{levis2019activity}
\bibinfo{author}{\bibfnamefont{D.}~\bibnamefont{Levis}}, \bibinfo{author}{\bibfnamefont{I.}~\bibnamefont{Pagonabarraga}}, \bibnamefont{and} \bibinfo{author}{\bibfnamefont{B.}~\bibnamefont{Liebchen}}, \bibinfo{journal}{Physical Review Research} \textbf{\bibinfo{volume}{1}}, \bibinfo{pages}{023026} (\bibinfo{year}{2019}).

\bibitem[{\citenamefont{Wang et~al.}(2015)\citenamefont{Wang, Duan, Ahmed, Sen, and Mallouk}}]{wang2015one}
\bibinfo{author}{\bibfnamefont{W.}~\bibnamefont{Wang}}, \bibinfo{author}{\bibfnamefont{W.}~\bibnamefont{Duan}}, \bibinfo{author}{\bibfnamefont{S.}~\bibnamefont{Ahmed}}, \bibinfo{author}{\bibfnamefont{A.}~\bibnamefont{Sen}}, \bibnamefont{and} \bibinfo{author}{\bibfnamefont{T.~E.} \bibnamefont{Mallouk}}, \bibinfo{journal}{Accounts of Chemical Research} \textbf{\bibinfo{volume}{48}}, \bibinfo{pages}{1938} (\bibinfo{year}{2015}).

\bibitem[{\citenamefont{Fern{\'a}ndez-Medina et~al.}(2020)\citenamefont{Fern{\'a}ndez-Medina, Ramos-Docampo, Hovorka, Salgueiri{\~n}o, and St{\"a}dler}}]{fernandez2020recent}
\bibinfo{author}{\bibfnamefont{M.}~\bibnamefont{Fern{\'a}ndez-Medina}}, \bibinfo{author}{\bibfnamefont{M.~A.} \bibnamefont{Ramos-Docampo}}, \bibinfo{author}{\bibfnamefont{O.}~\bibnamefont{Hovorka}}, \bibinfo{author}{\bibfnamefont{V.}~\bibnamefont{Salgueiri{\~n}o}}, \bibnamefont{and} \bibinfo{author}{\bibfnamefont{B.}~\bibnamefont{St{\"a}dler}}, \bibinfo{journal}{Advanced Functional Materials} \textbf{\bibinfo{volume}{30}}, \bibinfo{pages}{1908283} (\bibinfo{year}{2020}).

\bibitem[{\citenamefont{Yan et~al.}(2012)\citenamefont{Yan, Bloom, Bae, Luijten, and Granick}}]{yan2012linking}
\bibinfo{author}{\bibfnamefont{J.}~\bibnamefont{Yan}}, \bibinfo{author}{\bibfnamefont{M.}~\bibnamefont{Bloom}}, \bibinfo{author}{\bibfnamefont{S.~C.} \bibnamefont{Bae}}, \bibinfo{author}{\bibfnamefont{E.}~\bibnamefont{Luijten}}, \bibnamefont{and} \bibinfo{author}{\bibfnamefont{S.}~\bibnamefont{Granick}}, \bibinfo{journal}{Nature} \textbf{\bibinfo{volume}{491}}, \bibinfo{pages}{578} (\bibinfo{year}{2012}).

\bibitem[{\citenamefont{Urso et~al.}()\citenamefont{Urso, Ussia, and Pumera}}]{ursobreaking}
\bibinfo{author}{\bibfnamefont{M.}~\bibnamefont{Urso}}, \bibinfo{author}{\bibfnamefont{M.}~\bibnamefont{Ussia}}, \bibnamefont{and} \bibinfo{author}{\bibfnamefont{M.}~\bibnamefont{Pumera}}, \bibinfo{journal}{Advanced Functional Materials} p. \bibinfo{pages}{2101510} (????).

\bibitem[{\citenamefont{Dai et~al.}(2021)\citenamefont{Dai, Cheng, Li, Wang, Wang, Zheng, Liu, Chen, Wu, and Tang}}]{dai2021solution}
\bibinfo{author}{\bibfnamefont{J.}~\bibnamefont{Dai}}, \bibinfo{author}{\bibfnamefont{X.}~\bibnamefont{Cheng}}, \bibinfo{author}{\bibfnamefont{X.}~\bibnamefont{Li}}, \bibinfo{author}{\bibfnamefont{Z.}~\bibnamefont{Wang}}, \bibinfo{author}{\bibfnamefont{Y.}~\bibnamefont{Wang}}, \bibinfo{author}{\bibfnamefont{J.}~\bibnamefont{Zheng}}, \bibinfo{author}{\bibfnamefont{J.}~\bibnamefont{Liu}}, \bibinfo{author}{\bibfnamefont{J.}~\bibnamefont{Chen}}, \bibinfo{author}{\bibfnamefont{C.}~\bibnamefont{Wu}}, \bibnamefont{and} \bibinfo{author}{\bibfnamefont{J.}~\bibnamefont{Tang}}, \bibinfo{journal}{Advanced Functional Materials} p. \bibinfo{pages}{2106204} (\bibinfo{year}{2021}).

\bibitem[{\citenamefont{Vikrant and Kim}(2021)}]{vikrant2021metal}
\bibinfo{author}{\bibfnamefont{K.}~\bibnamefont{Vikrant}} \bibnamefont{and} \bibinfo{author}{\bibfnamefont{K.-H.} \bibnamefont{Kim}}, \bibinfo{journal}{Catalysis Science \& Technology}  (\bibinfo{year}{2021}).

\bibitem[{\citenamefont{Tesa{\v{r}} et~al.}(2022)\citenamefont{Tesa{\v{r}}, Ussia, Alduhaish, and Pumera}}]{tesavr2022autonomous}
\bibinfo{author}{\bibfnamefont{J.}~\bibnamefont{Tesa{\v{r}}}}, \bibinfo{author}{\bibfnamefont{M.}~\bibnamefont{Ussia}}, \bibinfo{author}{\bibfnamefont{O.}~\bibnamefont{Alduhaish}}, \bibnamefont{and} \bibinfo{author}{\bibfnamefont{M.}~\bibnamefont{Pumera}}, \bibinfo{journal}{Applied Materials Today} \textbf{\bibinfo{volume}{26}}, \bibinfo{pages}{101312} (\bibinfo{year}{2022}).

\bibitem[{\citenamefont{Li et~al.}(2015)\citenamefont{Li, Shklyaev, Li, Liu, Shum, Rozen, Balazs, and Wang}}]{li2015self}
\bibinfo{author}{\bibfnamefont{J.}~\bibnamefont{Li}}, \bibinfo{author}{\bibfnamefont{O.~E.} \bibnamefont{Shklyaev}}, \bibinfo{author}{\bibfnamefont{T.}~\bibnamefont{Li}}, \bibinfo{author}{\bibfnamefont{W.}~\bibnamefont{Liu}}, \bibinfo{author}{\bibfnamefont{H.}~\bibnamefont{Shum}}, \bibinfo{author}{\bibfnamefont{I.}~\bibnamefont{Rozen}}, \bibinfo{author}{\bibfnamefont{A.~C.} \bibnamefont{Balazs}}, \bibnamefont{and} \bibinfo{author}{\bibfnamefont{J.}~\bibnamefont{Wang}}, \bibinfo{journal}{Nano Letters} \textbf{\bibinfo{volume}{15}}, \bibinfo{pages}{7077} (\bibinfo{year}{2015}).

\bibitem[{\citenamefont{Cheng et~al.}(2014)\citenamefont{Cheng, Huang, Huang, Yang, Mao, Jin, ZhuGe, and Zhao}}]{cheng2014acceleration}
\bibinfo{author}{\bibfnamefont{R.}~\bibnamefont{Cheng}}, \bibinfo{author}{\bibfnamefont{W.}~\bibnamefont{Huang}}, \bibinfo{author}{\bibfnamefont{L.}~\bibnamefont{Huang}}, \bibinfo{author}{\bibfnamefont{B.}~\bibnamefont{Yang}}, \bibinfo{author}{\bibfnamefont{L.}~\bibnamefont{Mao}}, \bibinfo{author}{\bibfnamefont{K.}~\bibnamefont{Jin}}, \bibinfo{author}{\bibfnamefont{Q.}~\bibnamefont{ZhuGe}}, \bibnamefont{and} \bibinfo{author}{\bibfnamefont{Y.}~\bibnamefont{Zhao}}, \bibinfo{journal}{ACS Nano} \textbf{\bibinfo{volume}{8}}, \bibinfo{pages}{7746} (\bibinfo{year}{2014}).

\bibitem[{\citenamefont{Manamanchaiyaporn et~al.}(2021)\citenamefont{Manamanchaiyaporn, Tang, Yan, and Zheng}}]{manamanchaiyaporn2021molecular}
\bibinfo{author}{\bibfnamefont{L.}~\bibnamefont{Manamanchaiyaporn}}, \bibinfo{author}{\bibfnamefont{X.}~\bibnamefont{Tang}}, \bibinfo{author}{\bibfnamefont{X.}~\bibnamefont{Yan}}, \bibnamefont{and} \bibinfo{author}{\bibfnamefont{Y.}~\bibnamefont{Zheng}}, \bibinfo{journal}{IEEE Robotics and Automation Letters}  (\bibinfo{year}{2021}).

\bibitem[{\citenamefont{Linder and Halterman}(2014)}]{linder2014superconducting}
\bibinfo{author}{\bibfnamefont{J.}~\bibnamefont{Linder}} \bibnamefont{and} \bibinfo{author}{\bibfnamefont{K.}~\bibnamefont{Halterman}}, \bibinfo{journal}{Physical Review B} \textbf{\bibinfo{volume}{90}}, \bibinfo{pages}{104502} (\bibinfo{year}{2014}).

\bibitem[{\citenamefont{Iwasa and Tanaka}(2010)}]{iwasa2010dimensionality}
\bibinfo{author}{\bibfnamefont{M.}~\bibnamefont{Iwasa}} \bibnamefont{and} \bibinfo{author}{\bibfnamefont{D.}~\bibnamefont{Tanaka}}, \bibinfo{journal}{Physical Review E} \textbf{\bibinfo{volume}{81}}, \bibinfo{pages}{066214} (\bibinfo{year}{2010}).

\bibitem[{\citenamefont{Schilcher et~al.}(2021)\citenamefont{Schilcher, Schmidt, Vogell, and Bettstetter}}]{schilcher2021swarmalators}
\bibinfo{author}{\bibfnamefont{U.}~\bibnamefont{Schilcher}}, \bibinfo{author}{\bibfnamefont{J.~F.} \bibnamefont{Schmidt}}, \bibinfo{author}{\bibfnamefont{A.}~\bibnamefont{Vogell}}, \bibnamefont{and} \bibinfo{author}{\bibfnamefont{C.}~\bibnamefont{Bettstetter}}, in \emph{\bibinfo{booktitle}{2021 IEEE International Conference on Autonomic Computing and Self-Organizing Systems (ACSOS)}} (\bibinfo{organization}{IEEE}, \bibinfo{year}{2021}), pp. \bibinfo{pages}{90--99}.

\bibitem[{\citenamefont{Yadav et~al.}(2024)\citenamefont{Yadav, Chandrasekar, Zou, Kurths, and Senthilkumar}}]{yadav2024exotic}
\bibinfo{author}{\bibfnamefont{A.}~\bibnamefont{Yadav}}, \bibinfo{author}{\bibfnamefont{V.}~\bibnamefont{Chandrasekar}}, \bibinfo{author}{\bibfnamefont{W.}~\bibnamefont{Zou}}, \bibinfo{author}{\bibfnamefont{J.}~\bibnamefont{Kurths}}, \bibnamefont{and} \bibinfo{author}{\bibfnamefont{D.}~\bibnamefont{Senthilkumar}}, \bibinfo{journal}{Physical Review E} \textbf{\bibinfo{volume}{109}}, \bibinfo{pages}{044212} (\bibinfo{year}{2024}).

\bibitem[{\citenamefont{Lizarraga and de~Aguiar}(2020)}]{lizarraga2020synchronization}
\bibinfo{author}{\bibfnamefont{J.~U.} \bibnamefont{Lizarraga}} \bibnamefont{and} \bibinfo{author}{\bibfnamefont{M.~A.} \bibnamefont{de~Aguiar}}, \bibinfo{journal}{Chaos: An Interdisciplinary Journal of Nonlinear Science} \textbf{\bibinfo{volume}{30}}, \bibinfo{pages}{053112} (\bibinfo{year}{2020}).

\bibitem[{\citenamefont{Ha et~al.}(2019)\citenamefont{Ha, Jung, Kim, Park, and Zhang}}]{ha2019emergent}
\bibinfo{author}{\bibfnamefont{S.-Y.} \bibnamefont{Ha}}, \bibinfo{author}{\bibfnamefont{J.}~\bibnamefont{Jung}}, \bibinfo{author}{\bibfnamefont{J.}~\bibnamefont{Kim}}, \bibinfo{author}{\bibfnamefont{J.}~\bibnamefont{Park}}, \bibnamefont{and} \bibinfo{author}{\bibfnamefont{X.}~\bibnamefont{Zhang}}, \bibinfo{journal}{Mathematical Models and Methods in Applied Sciences} \textbf{\bibinfo{volume}{29}}, \bibinfo{pages}{2225} (\bibinfo{year}{2019}).

\bibitem[{\citenamefont{Ha et~al.}(2021)\citenamefont{Ha, Jung, Kim, Park, and Zhang}}]{ha2021mean}
\bibinfo{author}{\bibfnamefont{S.-Y.} \bibnamefont{Ha}}, \bibinfo{author}{\bibfnamefont{J.}~\bibnamefont{Jung}}, \bibinfo{author}{\bibfnamefont{J.}~\bibnamefont{Kim}}, \bibinfo{author}{\bibfnamefont{J.}~\bibnamefont{Park}}, \bibnamefont{and} \bibinfo{author}{\bibfnamefont{X.}~\bibnamefont{Zhang}}, \bibinfo{journal}{Kinetic \& Related Models}  (\bibinfo{year}{2021}).

\bibitem[{\citenamefont{O'Keeffe et~al.}(2018)\citenamefont{O'Keeffe, Evers, and Kolokolnikov}}]{o2018ring}
\bibinfo{author}{\bibfnamefont{K.~P.} \bibnamefont{O'Keeffe}}, \bibinfo{author}{\bibfnamefont{J.~H.} \bibnamefont{Evers}}, \bibnamefont{and} \bibinfo{author}{\bibfnamefont{T.}~\bibnamefont{Kolokolnikov}}, \bibinfo{journal}{Physical Review E} \textbf{\bibinfo{volume}{98}}, \bibinfo{pages}{022203} (\bibinfo{year}{2018}).

\bibitem[{\citenamefont{O'Keeffe et~al.}(2023)\citenamefont{O'Keeffe, Sar, Anwar, Liz{\'a}rraga, de~Aguiar, and Ghosh}}]{o2023solvable}
\bibinfo{author}{\bibfnamefont{K.}~\bibnamefont{O'Keeffe}}, \bibinfo{author}{\bibfnamefont{G.~K.} \bibnamefont{Sar}}, \bibinfo{author}{\bibfnamefont{M.~S.} \bibnamefont{Anwar}}, \bibinfo{author}{\bibfnamefont{J.~U.} \bibnamefont{Liz{\'a}rraga}}, \bibinfo{author}{\bibfnamefont{M.~A.} \bibnamefont{de~Aguiar}}, \bibnamefont{and} \bibinfo{author}{\bibfnamefont{D.}~\bibnamefont{Ghosh}}, \bibinfo{journal}{arXiv preprint arXiv:2312.10178}  (\bibinfo{year}{2023}).

\bibitem[{\citenamefont{Liz{\'a}rraga et~al.}(2024)\citenamefont{Liz{\'a}rraga, O'Keeffe, and de~Aguiar}}]{lizarraga2024order}
\bibinfo{author}{\bibfnamefont{J.~U.} \bibnamefont{Liz{\'a}rraga}}, \bibinfo{author}{\bibfnamefont{K.~P.} \bibnamefont{O'Keeffe}}, \bibnamefont{and} \bibinfo{author}{\bibfnamefont{M.~A.} \bibnamefont{de~Aguiar}}, \bibinfo{journal}{Physical Review E} \textbf{\bibinfo{volume}{109}}, \bibinfo{pages}{044209} (\bibinfo{year}{2024}).

\bibitem[{\citenamefont{O'Keeffe et~al.}(2022)\citenamefont{O'Keeffe, Ceron, and Petersen}}]{o2022collective}
\bibinfo{author}{\bibfnamefont{K.}~\bibnamefont{O'Keeffe}}, \bibinfo{author}{\bibfnamefont{S.}~\bibnamefont{Ceron}}, \bibnamefont{and} \bibinfo{author}{\bibfnamefont{K.}~\bibnamefont{Petersen}}, \bibinfo{journal}{Physical Review E} \textbf{\bibinfo{volume}{105}}, \bibinfo{pages}{014211} (\bibinfo{year}{2022}).

\bibitem[{\citenamefont{Yoon et~al.}(2022)\citenamefont{Yoon, O’Keeffe, Mendes, and Goltsev}}]{yoon2022sync}
\bibinfo{author}{\bibfnamefont{S.}~\bibnamefont{Yoon}}, \bibinfo{author}{\bibfnamefont{K.}~\bibnamefont{O’Keeffe}}, \bibinfo{author}{\bibfnamefont{J.}~\bibnamefont{Mendes}}, \bibnamefont{and} \bibinfo{author}{\bibfnamefont{A.}~\bibnamefont{Goltsev}}, \bibinfo{journal}{Physical Review Letters} \textbf{\bibinfo{volume}{129}}, \bibinfo{pages}{208002} (\bibinfo{year}{2022}).

\bibitem[{\citenamefont{Hong et~al.}(2023)\citenamefont{Hong, O'Keeffe, Lee, and Park}}]{hong2023swarmalators}
\bibinfo{author}{\bibfnamefont{H.}~\bibnamefont{Hong}}, \bibinfo{author}{\bibfnamefont{K.~P.} \bibnamefont{O'Keeffe}}, \bibinfo{author}{\bibfnamefont{J.~S.} \bibnamefont{Lee}}, \bibnamefont{and} \bibinfo{author}{\bibfnamefont{H.}~\bibnamefont{Park}}, \bibinfo{journal}{Physical Review Research} \textbf{\bibinfo{volume}{5}}, \bibinfo{pages}{023105} (\bibinfo{year}{2023}).

\bibitem[{\citenamefont{Anwar et~al.}(2024{\natexlab{a}})\citenamefont{Anwar, Sar, Perc, and Ghosh}}]{anwar2024collective}
\bibinfo{author}{\bibfnamefont{M.~S.} \bibnamefont{Anwar}}, \bibinfo{author}{\bibfnamefont{G.~K.} \bibnamefont{Sar}}, \bibinfo{author}{\bibfnamefont{M.}~\bibnamefont{Perc}}, \bibnamefont{and} \bibinfo{author}{\bibfnamefont{D.}~\bibnamefont{Ghosh}}, \bibinfo{journal}{Communications Physics} \textbf{\bibinfo{volume}{7}}, \bibinfo{pages}{59} (\bibinfo{year}{2024}{\natexlab{a}}).

\bibitem[{\citenamefont{Liz{\'a}rraga and de~Aguiar}(2023)}]{lizarraga2023synchronization}
\bibinfo{author}{\bibfnamefont{J.~U.} \bibnamefont{Liz{\'a}rraga}} \bibnamefont{and} \bibinfo{author}{\bibfnamefont{M.~A.} \bibnamefont{de~Aguiar}}, \bibinfo{journal}{Physical Review E} \textbf{\bibinfo{volume}{108}}, \bibinfo{pages}{024212} (\bibinfo{year}{2023}).

\bibitem[{\citenamefont{Sar et~al.}(2023{\natexlab{a}})\citenamefont{Sar, O’Keeffe, and Ghosh}}]{sar2023swarmalators}
\bibinfo{author}{\bibfnamefont{G.~K.} \bibnamefont{Sar}}, \bibinfo{author}{\bibfnamefont{K.}~\bibnamefont{O’Keeffe}}, \bibnamefont{and} \bibinfo{author}{\bibfnamefont{D.}~\bibnamefont{Ghosh}}, \bibinfo{journal}{Chaos: An Interdisciplinary Journal of Nonlinear Science} \textbf{\bibinfo{volume}{33}}, \bibinfo{pages}{111103} (\bibinfo{year}{2023}{\natexlab{a}}).

\bibitem[{\citenamefont{O'Keeffe and Hong}(2022)}]{o2022swarmalators}
\bibinfo{author}{\bibfnamefont{K.}~\bibnamefont{O'Keeffe}} \bibnamefont{and} \bibinfo{author}{\bibfnamefont{H.}~\bibnamefont{Hong}}, \bibinfo{journal}{Physical Review E} \textbf{\bibinfo{volume}{105}}, \bibinfo{pages}{064208} (\bibinfo{year}{2022}).

\bibitem[{\citenamefont{Hao et~al.}(2023)\citenamefont{Hao, Zhong, and O'Keeffe}}]{hao2023attractive}
\bibinfo{author}{\bibfnamefont{B.}~\bibnamefont{Hao}}, \bibinfo{author}{\bibfnamefont{M.}~\bibnamefont{Zhong}}, \bibnamefont{and} \bibinfo{author}{\bibfnamefont{K.}~\bibnamefont{O'Keeffe}}, \bibinfo{journal}{Physical Review E} \textbf{\bibinfo{volume}{108}}, \bibinfo{pages}{064214} (\bibinfo{year}{2023}).

\bibitem[{\citenamefont{Anwar et~al.}(2024{\natexlab{b}})\citenamefont{Anwar, Ghosh, and O'Keeffe}}]{anwar2024forced}
\bibinfo{author}{\bibfnamefont{M.~S.} \bibnamefont{Anwar}}, \bibinfo{author}{\bibfnamefont{D.}~\bibnamefont{Ghosh}}, \bibnamefont{and} \bibinfo{author}{\bibfnamefont{K.}~\bibnamefont{O'Keeffe}}, \bibinfo{journal}{arXiv preprint arXiv:2409.05342}  (\bibinfo{year}{2024}{\natexlab{b}}).

\bibitem[{\citenamefont{O'Keeffe}(2024)}]{o2024stability}
\bibinfo{author}{\bibfnamefont{K.}~\bibnamefont{O'Keeffe}}, \bibinfo{journal}{arXiv preprint arXiv:2410.02975}  (\bibinfo{year}{2024}).

\bibitem[{\citenamefont{Lu and Steinerberger}(2020)}]{lu2020synchronization}
\bibinfo{author}{\bibfnamefont{J.}~\bibnamefont{Lu}} \bibnamefont{and} \bibinfo{author}{\bibfnamefont{S.}~\bibnamefont{Steinerberger}}, \bibinfo{journal}{Nonlinearity} \textbf{\bibinfo{volume}{33}}, \bibinfo{pages}{5905} (\bibinfo{year}{2020}).

\bibitem[{\citenamefont{Ling et~al.}(2019)\citenamefont{Ling, Xu, and Bandeira}}]{ling2019landscape}
\bibinfo{author}{\bibfnamefont{S.}~\bibnamefont{Ling}}, \bibinfo{author}{\bibfnamefont{R.}~\bibnamefont{Xu}}, \bibnamefont{and} \bibinfo{author}{\bibfnamefont{A.~S.} \bibnamefont{Bandeira}}, \bibinfo{journal}{SIAM Journal on Optimization} \textbf{\bibinfo{volume}{29}}, \bibinfo{pages}{1879} (\bibinfo{year}{2019}).

\bibitem[{\citenamefont{Taylor}(2012)}]{taylor2012there}
\bibinfo{author}{\bibfnamefont{R.}~\bibnamefont{Taylor}}, \bibinfo{journal}{Journal of Physics A: Mathematical and Theoretical} \textbf{\bibinfo{volume}{45}}, \bibinfo{pages}{055102} (\bibinfo{year}{2012}).

\bibitem[{\citenamefont{Kassabov et~al.}(2022)\citenamefont{Kassabov, Strogatz, and Townsend}}]{kassabov2022global}
\bibinfo{author}{\bibfnamefont{M.}~\bibnamefont{Kassabov}}, \bibinfo{author}{\bibfnamefont{S.~H.} \bibnamefont{Strogatz}}, \bibnamefont{and} \bibinfo{author}{\bibfnamefont{A.}~\bibnamefont{Townsend}}, \bibinfo{journal}{Chaos: An Interdisciplinary Journal of Nonlinear Science} \textbf{\bibinfo{volume}{32}} (\bibinfo{year}{2022}).

\bibitem[{\citenamefont{Abdalla et~al.}(2024)\citenamefont{Abdalla, Bandeira, and Invernizzi}}]{abdalla2024guarantees}
\bibinfo{author}{\bibfnamefont{P.}~\bibnamefont{Abdalla}}, \bibinfo{author}{\bibfnamefont{A.~S.} \bibnamefont{Bandeira}}, \bibnamefont{and} \bibinfo{author}{\bibfnamefont{C.}~\bibnamefont{Invernizzi}}, \bibinfo{journal}{SIAM Journal on Applied Dynamical Systems} \textbf{\bibinfo{volume}{23}}, \bibinfo{pages}{779} (\bibinfo{year}{2024}).

\bibitem[{\citenamefont{Nagpal et~al.}(2024)\citenamefont{Nagpal, Nair, Strogatz, and Parise}}]{nagpal2024synchronization}
\bibinfo{author}{\bibfnamefont{S.~V.} \bibnamefont{Nagpal}}, \bibinfo{author}{\bibfnamefont{G.~G.} \bibnamefont{Nair}}, \bibinfo{author}{\bibfnamefont{S.~H.} \bibnamefont{Strogatz}}, \bibnamefont{and} \bibinfo{author}{\bibfnamefont{F.}~\bibnamefont{Parise}}, \bibinfo{journal}{arXiv preprint arXiv:2403.13998}  (\bibinfo{year}{2024}).

\bibitem[{\citenamefont{Harrington et~al.}(2023)\citenamefont{Harrington, Schenck, and Stillman}}]{harrington2023kuramoto}
\bibinfo{author}{\bibfnamefont{H.}~\bibnamefont{Harrington}}, \bibinfo{author}{\bibfnamefont{H.}~\bibnamefont{Schenck}}, \bibnamefont{and} \bibinfo{author}{\bibfnamefont{M.}~\bibnamefont{Stillman}}, \bibinfo{journal}{arXiv preprint arXiv:2312.16069}  (\bibinfo{year}{2023}).

\bibitem[{\citenamefont{Sar et~al.}(2023{\natexlab{b}})\citenamefont{Sar, Ghosh, and O'Keeffe}}]{sar2023pinning}
\bibinfo{author}{\bibfnamefont{G.~K.} \bibnamefont{Sar}}, \bibinfo{author}{\bibfnamefont{D.}~\bibnamefont{Ghosh}}, \bibnamefont{and} \bibinfo{author}{\bibfnamefont{K.}~\bibnamefont{O'Keeffe}}, \bibinfo{journal}{Physical Review E} \textbf{\bibinfo{volume}{107}}, \bibinfo{pages}{024215} (\bibinfo{year}{2023}{\natexlab{b}}).

\bibitem[{\citenamefont{Pedergnana and Noiray}(2022)}]{pedergnana2022exact}
\bibinfo{author}{\bibfnamefont{T.}~\bibnamefont{Pedergnana}} \bibnamefont{and} \bibinfo{author}{\bibfnamefont{N.}~\bibnamefont{Noiray}}, \bibinfo{journal}{Chaos: An Interdisciplinary Journal of Nonlinear Science} \textbf{\bibinfo{volume}{32}} (\bibinfo{year}{2022}).

\bibitem[{\citenamefont{Pedergnana and Noiray}(2023)}]{pedergnana2023certain}
\bibinfo{author}{\bibfnamefont{T.}~\bibnamefont{Pedergnana}} \bibnamefont{and} \bibinfo{author}{\bibfnamefont{N.}~\bibnamefont{Noiray}}, \bibinfo{journal}{arXiv preprint arXiv:2309.02513}  (\bibinfo{year}{2023}).

\bibitem[{Note1()}]{Note1}
Note1, \bibinfo{note}{by rotational symmetry we mean you can add a constant to each phase $\xi _i, \eta _i \rightarrow \theta _i + C_a, C_b$ for constants $C_a, C_b$ and the dynamics don't change (because only phase differences $\xi _j - \xi _i$ etc appear in the governing ODEs)}.

\bibitem[{\citenamefont{Khev}(2024{\natexlab{a}})}]{github}
\bibinfo{author}{\bibnamefont{Khev}}, \emph{\bibinfo{title}{Swarmalators: 1d on ring}} (\bibinfo{year}{2024}{\natexlab{a}}), \bibinfo{note}{accessed: 2024-09-24}, \urlprefix\url{https://github.com/Khev/swarmalators/tree/master/1D/on-ring/global-stability}.

\bibitem[{\citenamefont{Watanabe and Strogatz}(1993)}]{watanabe1993integrability}
\bibinfo{author}{\bibfnamefont{S.}~\bibnamefont{Watanabe}} \bibnamefont{and} \bibinfo{author}{\bibfnamefont{S.~H.} \bibnamefont{Strogatz}}, \bibinfo{journal}{Physical review letters} \textbf{\bibinfo{volume}{70}}, \bibinfo{pages}{2391} (\bibinfo{year}{1993}).

\bibitem[{\citenamefont{Khev}(2024{\natexlab{b}})}]{khev2024swarmalators}
\bibinfo{author}{\bibnamefont{Khev}}, \emph{\bibinfo{title}{Swarmalators: 1d on-ring global stability}}, \bibinfo{howpublished}{\url{https://github.com/Khev/swarmalators/tree/master/1D/on-ring/global-stability}} (\bibinfo{year}{2024}{\natexlab{b}}), \bibinfo{note}{accessed: 2024-09-12}.

\bibitem[{\citenamefont{Barci{\'s} and Bettstetter}(2020)}]{barcis2020sandsbots}
\bibinfo{author}{\bibfnamefont{A.}~\bibnamefont{Barci{\'s}}} \bibnamefont{and} \bibinfo{author}{\bibfnamefont{C.}~\bibnamefont{Bettstetter}}, \bibinfo{journal}{IEEE Access} \textbf{\bibinfo{volume}{8}}, \bibinfo{pages}{218752} (\bibinfo{year}{2020}).

\bibitem[{\citenamefont{Yeung and Strogatz}(1999)}]{yeung1999time}
\bibinfo{author}{\bibfnamefont{M.~S.} \bibnamefont{Yeung}} \bibnamefont{and} \bibinfo{author}{\bibfnamefont{S.~H.} \bibnamefont{Strogatz}}, \bibinfo{journal}{Physical Review Letters} \textbf{\bibinfo{volume}{82}}, \bibinfo{pages}{648} (\bibinfo{year}{1999}).

\bibitem[{\citenamefont{Blum et~al.}(2024)\citenamefont{Blum, Li, O'Keeffe, and Kogan}}]{blum2024swarmalators}
\bibinfo{author}{\bibfnamefont{N.}~\bibnamefont{Blum}}, \bibinfo{author}{\bibfnamefont{A.}~\bibnamefont{Li}}, \bibinfo{author}{\bibfnamefont{K.}~\bibnamefont{O'Keeffe}}, \bibnamefont{and} \bibinfo{author}{\bibfnamefont{O.}~\bibnamefont{Kogan}}, \bibinfo{journal}{Physical Review E} \textbf{\bibinfo{volume}{109}}, \bibinfo{pages}{014205} (\bibinfo{year}{2024}).

\bibitem[{Note2()}]{Note2}
Note2, \bibinfo{note}{with $\delta t$ large enough that the $\delta t \rightarrow 0$ limit is inappropriate, and so we can't use ODE approximations}.

\bibitem[{\citenamefont{Peskin}(1975)}]{peskin1975mathematical}
\bibinfo{author}{\bibfnamefont{C.~S.} \bibnamefont{Peskin}}, \bibinfo{journal}{Courant Inst. Math}  (\bibinfo{year}{1975}).

\bibitem[{\citenamefont{Mirollo and Strogatz}(1990)}]{mirollo1990synchronization}
\bibinfo{author}{\bibfnamefont{R.~E.} \bibnamefont{Mirollo}} \bibnamefont{and} \bibinfo{author}{\bibfnamefont{S.~H.} \bibnamefont{Strogatz}}, \bibinfo{journal}{SIAM Journal on Applied Mathematics} \textbf{\bibinfo{volume}{50}}, \bibinfo{pages}{1645} (\bibinfo{year}{1990}).

\bibitem[{\citenamefont{O’Keeffe et~al.}(2015)\citenamefont{O’Keeffe, Krapivsky, and Strogatz}}]{o2015synchronization}
\bibinfo{author}{\bibfnamefont{K.~P.} \bibnamefont{O’Keeffe}}, \bibinfo{author}{\bibfnamefont{P.~L.} \bibnamefont{Krapivsky}}, \bibnamefont{and} \bibinfo{author}{\bibfnamefont{S.~H.} \bibnamefont{Strogatz}}, \bibinfo{journal}{Physical review letters} \textbf{\bibinfo{volume}{115}}, \bibinfo{pages}{064101} (\bibinfo{year}{2015}).

\bibitem[{\citenamefont{O'Keeffe}(2016)}]{o2016transient}
\bibinfo{author}{\bibfnamefont{K.~P.} \bibnamefont{O'Keeffe}}, \bibinfo{journal}{Physical Review E} \textbf{\bibinfo{volume}{93}}, \bibinfo{pages}{032203} (\bibinfo{year}{2016}).

\bibitem[{\citenamefont{Gwon and Cho}(2023)}]{gwon2023percolation}
\bibinfo{author}{\bibfnamefont{G.}~\bibnamefont{Gwon}} \bibnamefont{and} \bibinfo{author}{\bibfnamefont{Y.~S.} \bibnamefont{Cho}}, \bibinfo{journal}{Chaos: An Interdisciplinary Journal of Nonlinear Science} \textbf{\bibinfo{volume}{33}} (\bibinfo{year}{2023}).

\end{thebibliography}

\end{document}


\begin{lemma}[Stability of $\pi_{p0}, \pi_{0p}$ states]
The $\pi_{p0}$ states are those where $p$ swarmalators have $\xi_i=C$ and the remaining $N-p$ have $\xi = C+\pi$. Their $\eta_i$ are fully desync $s = 0$. The $\pi_{0p}$ states are the same with the roles of $\xi,\eta$ swapped. Both states are unstable on $D_{sync}$.
\end{lemma}

\begin{proof}
This proof was the trickiest. We had to graph. First we stared with

The characteristic equation can be factored into form \cite{github}
\begin{align}
\lambda^{N-1} \Big( a + b \lambda + c \lambda^2 \Big) \Big( a + b \lambda + c \lambda^2 \Big)
\end{align}
The Routh-Hurwitz conditions require $a_3 > 0$, which we can see is not true when $K>0$ which is in the sync region. Thus, the states are all unstable on $D_{sync}$.
\end{proof}

\textit{Remark}. We conjecture that the $\pi_{p0}, \pi_{0p}$ states are unstable for \textit{all} parameter values; proving this is an open problem.

xxx

\section{Linearly spaced}
We find the eigenvalues as before. Table~\ref{take} shows that there are a small number of unique ones that depend on $p$; it saturates at 10 for $p > 3$. Recall that we require $p \leq N/2$. Table~\ref{lambda_vals_flipped} gives a numerical examples for $N=3$ and $(J,K) = (1,0.21)$.

\begin{table}[h!]
\centering
\begin{tabular}{@{}cccccccccc@{}}
\toprule
$p$                & 1  & 2  & 3  & 4  & 5  & \dots & $n/2$ \\ \midrule
\# distinct $\lambda$ & 7  & 8  & 9  & 10 & 10 & \dots & 10    \\ \bottomrule
\end{tabular}
\caption{Number of distinct $\lambda$ values for different $p$.} \label{take}
\end{table}

\begin{table}[h!]
\centering
\small 
\begin{tabular}{@{}ccccccc@{}}
\toprule
$\lambda$ & $p = 1$ & $p = 2$ & $p = 3$ & $p = 4$ & $p = 5$ & $p = 6$ \\ \midrule
$\lambda_1$ & -0.18 & -0.15 & -0.19 - 0.23i & -0.23 - 0.03i & -0.43 & -0.48 \\
$\lambda_2$ & -0.08 - 0.53i & -0.14 - 0.37i & -0.19 + 0.23i & -0.23 + 0.03i & -0.07 & -0.02 \\
$\lambda_3$ & -0.08 + 0.53i & -0.14 + 0.37i & -0.11 & -0.08 & -0.05 & -0.02 \\
$\lambda_4$ & -0.04 - 0.63i & -0.02 - 0.56i & -0.01 - 0.47i & -0.01 - 0.36i & -0.00 - 0.22i & 0.00 \\
$\lambda_5$ & -0.04 + 0.63i & -0.02 + 0.56i & -0.01 + 0.47i & -0.01 + 0.36i & -0.00 + 0.22i & 0.00 - 0.08i \\
$\lambda_6$ & 0.00 & 0.00 & 0.00 & 0.00 & 0.00 & 0.00 + 0.08i \\
$\lambda_7$ & 0.30 & 0.15 & 0.11 & 0.08 & 0.05 & 0.02 \\
$\lambda_8$ & & 0.45 & 0.13 & 0.08 & 0.05 & 0.02 \\
$\lambda_9$ & & & 0.59 & 0.12 & 0.11 & 0.10 \\
$\lambda_{10}$ & & & & 0.70 & 0.77 & 0.80 \\ \bottomrule
\end{tabular}
\caption{Numerically computed eigenvalues $\lambda$ for different $p$. Here $N = 13$ and $(J,K) = (1,0.21)$. Duplicate values have been omitted.} \label{lambda_vals_flipped}
\end{table}

Moreover, the characteristic equations can be factored into polynomials whose stability properties can be determined using the Routh-Hurwitz conditions. We do not use all of the conditions—only the first one is required. 

For a polynomial of degree $n$
\begin{align}
a_n \lambda^n + a_{n-1} \lambda^{n-1} + \dots + a_0 = 0,
\end{align}
the Routh-Hurwitz conditions tell us
\begin{align}
    a_{n}, a_{n-1} > 0 \label{cond}
\end{align}
are necessary (but not sufficient) for stability. We will use these to rule \textit{out} stability for all $J,K$ of the $\pi_{p0}$ states. We take each $p$ case by case. 

For $p=1$, the characteristic equation factors into a product of polynomials of degree 1,2,3. 
\begin{align}
\lambda^{N-1}(a_{0,n} + a_{1,n} \lambda)^{N-4}(b_{0,n} + K b_{1,n} \lambda + b_{2,n} \lambda^2)( c_{0,n} + c_{1,n} \lambda  -K c_{2,n} \lambda^2 + \lambda^3  ) = 0
\end{align}
\noindent

The coefficients are functions of $N,J,K$ and can have either sign. The exception is $c_{2,n}>0$ and $b_{2,n}>0$ which are strictly positive. Recalling the stability conditions Eq.~\eqref{cond}, we see the second order polynomial requires $K>0$ while the third order one requires $K<0$ -- thus this is unstable for all $K$. 

For $p=2$, the characteristic equation factors into a product like so
\begin{align}
\lambda^{N-1}(a_{0,n} + a_{1,n} \lambda)^{N-5}( \lambda^6 + K c_{n,5} \lambda^5 + \dots ) = 0
\end{align}
\noindent
where $a_{0,n}, a_{1,n}, c_{n,5} > 0$. The first term requires $K>0$, while the second sixth order requires $K<0$, which again implies instability for all $K$.

For $p=3$, the characteristic equation factors into a product like so
\begin{align}
\lambda^{N-1}(a_{0,n} + a_{1,n} \lambda)^{N-6}( \lambda^7 + K c_{n,6} \lambda^6 + \dots ) = 0
\end{align}
\noindent
where $a_{0,n}, a_{1,n}, c_{n,6} > 0$. The story is the same as above: the first term requires $K>0$, while the second sixth order requires $K<0$, which again implies instability for all $K$.

For $3<p<N/2$, all that changes is the order of the polynomial increases by one
\begin{align}
\lambda^{N-1}(a_{0,n} + a_{1,n} \lambda)^{N-7}( \lambda^8 + K c_{n,7} \lambda^7 + \dots ) = 0
\end{align}
\noindent
but the instability for all $K$ holds. This proves our result.

\section{General result}
Numerics indicate that these states are unstable for all the parameter values. Proving this was tricky because the characteristic equations for the eigenvalues $\lambda$ had both an $N$ and $p$ dependence. However after much experimentation, patterns emerged. We found two classes of characteristic equations separated by a threshold boundary $p^*(N)$
\begin{align}
& \lambda^{N-1} \Big( a_0 + a_1 \lambda + \dots - \mu(N,p) K \lambda^{N} + \lambda^{N+1} \Big) = 0 , \quad \text{for } p \leq p^* \label{f1} \\
& \lambda^{N-1} \Big( \lambda + C(N,p) K \Big)^{N-m} \Big( b_0 + b_1 \lambda + \lambda^2 \Big) \times \\ & \hspace{0.85 cm} \Big( c_0 + c_1 \lambda + c_2 \lambda^2 - \nu(N,p) K \lambda^{m-1} + \lambda^m \Big) = 0, \quad \text{for } p > p^* \label{f2}
\end{align}
with 
\begin{align}
    p^*(N) &= \frac{1}{2}(N - \sqrt{2 N}) \\ 
    \mu(N,p) &=  N - 2 - 4p + \frac{4p^2}{N} \\
    C(N,p) &= \frac{1}{2} \Big(1 - \frac{2p}{N} \Big) \\
    \nu(N,p) &= \mu(N,p) - (N-m) C(N,p) 
\end{align}
where $m$ is an exponent that depends on $p$: $(p,m) = (1,4), (2,5), (3,6)$ and for $p>3$ $N=6$. In Eq~\ref{f1}, the coefficient $a_i=a_i(J,K, \xi_1, \dots \xi_2)$ are complex expressions whose form is unimportant. All that matters is the $\mu(N,p)$ term which is strictly positive $\mu(N,p) > 0$ for $p \leq p*$ (we will derive it shortly). Recall a necessary condition for stability (from the Routh Hurwitz conditions) is the $n-1$-th term $c_{n-1} > 0$. For our problem, since $\mu(N,p) >0$ this requires $K<0$. Thus, the $\pi_{p0}$ states are \textit{unstable} on $K>0$ which constains the sync region $D_{sync}$. (We believe they are unstable for all $K$, but we don't need to prove that here). The case $p > p^*(N)$, is the same. You can see the $n-1$-th term in the third polynomial is positive only when $K<0$ proving instability on $K>0$ and in turn $D_{sync}$. Taken together, these show the $\pi_{p0}$ states are unstable on $D_{sync}$ as claimed.

Now we derive Eq.~\eqref{f1}. Straightforward algebra shows the $\lambda^{N-1}$ pre factor pops out, leaving a polynomial of degree $N-1$ in $\lambda$. We need to show the second to last term is $\mu(N,p)$. Recall this second to last term is minus the trace of the matrix $M$ which generated the characteristic polynomial. 
\begin{align}
    \mu(N,p) = - Tr(M)
\end{align}
Since $M$ is block diagonal, $Tr(M) = k Tr(M_1) + k Tr(M_2)$. $Tr(M_1) = 1$. $M_2$ takes a little more work, but comes out as $x=1$. Putting this together gives
\begin{align}
    \mu_N = -K \left( N - 2 - 4p + \frac{4p^2}{N} \right)
\end{align}
The point at which this changes sign (solving $\mu_N = 0$) gives us the critical boundary $p^*(N) = \frac{1}{2}( N - \sqrt{2 N})$ reported above. Thus for $p < p^*$, $\mu_{N,p} > 0$ as required. Figure~\ref{} checked out calculation numerically by computing the sign of $\mu_N$ over a $(p,N)$ mesh. Blue regions are where $\mu_{N}$, red when it is negative. As you can see, the jagged line.

Final task is Eq.~\eqref{f2}. This one we did by trial and error. We found a root $\lambda = \beta_{N,p}$ and verified it was valid via substitution. Factoring this out then led to the desired equation.

\subsection{p=1}
The cases $N=3,4$ are special. The char equations in this case take for
\begin{align}
    \lambda^{N-1} \Big( a_0 + a_1 \lambda + a_2 \lambda^2  - \mu_{N} K \lambda^3 + \lambda^4  \Big)
\end{align}
where $\mu(N=3) = 5/3$ and $ \mu(N=4) = 1$. The RH condition $a_{3} > 0 $ tell us both states are unstable when $K>0$ which is the sync region. (We believe they are unstable for \textit{all}, but proving that is unnecessary here).

For $N>4$, the characteristic equation takes form
\begin{align}
\lambda^{N-1} \Big( \lambda + \frac{N-2}{N} K \Big)^{N-4} \Big( a_0 + \frac{n-4}{2n} K \lambda + \lambda^2 \Big) \Big( b_0 + b_1 \lambda + b_2 \lambda^2 - \frac{n+4}{2n} K \lambda^3 + \lambda^4 \Big)
\end{align}
where the $a_0, b_0, b_1, b_2$ are complex function involving strings of $\cos \xi$. But their values don't matter. We can see that the root $\lambda = - K(N-2)/N$ requires $K>0$ to be positive, but the requirement of $b_3>0$ requires $K<0$, meaning the state is unstable for all $K$. 

(MAYBE ADD IN NUMERICS SHOWING THE ROOT IS THERE).